\documentstyle[eqsecnum,aps,epsf,twocolumn,]{revtex}

\input amssym.def
\input amssym.tex

\newcommand{\vc}[1]{\mbox{\bm$#1$}}

\newcommand{\li}[2]{\mbox{${\rm Li}_{\scriptscriptstyle{#1}}\left[#2\right]$}}
\newcommand{\zer}[1]{\mbox{$\zeta_{\scriptscriptstyle R}\left(#1\right)$}}
\newcommand{\zertwo}[1]{\mbox{$\zeta^{2}_{\scriptscriptstyle R}\left(#1\right)$}}
\newcommand{\zerthree}[1]{\mbox{$\zeta^{3}_{\scriptscriptstyle R}\left(#1\right)$}}
\newcommand{\zeh}[2]{\mbox{$\zeta_{\scriptscriptstyle H}\left(#1,#2\right)$}}
\newcommand{\zehtwo}[2]{\mbox{$\zeta^{2}_{\scriptscriptstyle H}\left(#1,#2\right)$}}
\newcommand{\bes}[3]{\mbox{$\Sigma_{\sss{#1}}\left[#2,#3\right]$}}
\newcommand{\bm}{\boldmath}
\newcommand{\sss}{\scriptscriptstyle}
\newcommand{\sstack}[2]{\stackrel{\scriptstyle{#1}}{#2}}
\newcommand{\nn}{\nonumber}
\newcommand{\Real}{{\Re}e\;}

\begin{document}
\preprint{}
\bibliographystyle{prsty}

\title{Bose--Einstein condensation of nonrelativistic charged particles
in a constant magnetic field} 
\author{Guy B. Standen and David J. Toms}
\address{Physics Department, University of Newcastle, UK, NE1 7RU}
\date{\today}
\maketitle 
\begin{abstract}
The statistical mechanics of a system of non--relativistic charged particles in a constant magnetic field
is discussed. The spatial dimension $D$ is arbitrary with $D\geq 3$ assumed. Calculations are presented
from first principles using the effective action method. For $D\geq 5$ the system has a phase transition
with a Bose condensate. We show how the effective action method method deals with in a very natural way
with the condensate and study it's r\^{o}le in the magnetization of the gas. For large values of the magnetic 
field  we show how the magnetized gas in $D$ spatial dimensions behaves like the free Bose gas in $(D-2)$
spatial dimensions. Even though for $D=3$ the magnetized gas does not have a phase transition for any 
non--zero value of the magnetic field, we show how the specific heat starts to resemble the result for the
free gas as the magnetic field is reduced. A number of analytical approximations for the magnetization and 
specific heat are given and compared with numerical results. In this way we are able to study in precise 
detail how the $B\rightarrow 0$ limit of the magnetized gas is achieved.
\end{abstract}
\pacs{03.75.Fi, 05.30.Jp}

\narrowtext

\section{Introduction}
\label{sec:intro}

Bose-Einstein condensation (BEC) \cite{bose24a,eins24a} has been an
integral part of physics for many years. From a theoretical standpoint this phenomenon 
has been intensively studied as a possible means of explaining the superfluid 
transition in He$^{4}$ \cite{lond38a}, and also as an experimentally realisable possibility
in dilute atomic gases. Recently BEC has been achieved in atomic systems of dilute 
alkali metal atoms 
\cite{andeenshmattwiemcorn95a,bradsacktollhule95a,davimeweandrdrutdurfkurnkett95a,meweandrdrutkurndurfkett96a}, 
and for this reason a great deal of interest is now being focussed on these systems.

Another system worthy of study is the condensation (or lack thereof) of
charged bosons in a homogeneous magnetic field. Initial work on the
nonrelativistic case \cite{osbo49b} was substantially improved on by
Schafroth \cite{scha55a}, and was then generalized to dimensions other than three
by May \cite{may59a,may65a}. The primary motivation for these efforts
was to formulate a plausible theory of superconductivity. Although this
attempt failed as far as standard low T$_c$ superconductors are
concerned, interest in the subject has revived as explanations of high
T$_c$ materials are now required \cite{hirsscal85a}. 

More recently, several detailed studies have been made of the ideal system 
of nonrelativistic charged bosons, both numerically \cite{ariajoan89a}, and 
from a theoretical point of view in both two \cite{daicfran97a}, 
and three \cite{daicfrangailkowa94a,toms94a,daicfran96a,roja96a} dimensions. 
Also a detailed study of the boson gas in three dimensions with trapping harmonic potentials
and the presence of a magnetic field has recently been made using the path--integral formalism 
\cite{brosdevrlemm97a}. These authors find comparable behavior to
that found using Mellin transforms, and a comparison between the two methods is very instructive.
However due to constraints of space, the contrast between the two methods will not be discussed
here in depth.

The behavior of the relativistic case has also been discussed 
\cite{millray86a,toms94b,elmfliljpersskag95,daicfran96a}. 
In addition more realistic systems which may occur in high T$_{c}$ 
superconductors \cite{alexrubi93a,alex93a,alexbeerkaba96a} have also been examined.

The outline of our paper is as follows. In section \ref{sec:becnob} we give a brief description of the 
free Bose gas in $D$ spatial dimensions. We will later compare the analogous results for the magnetized 
gas to these free--field ones. Section \ref{sec:becwib} presents the effective action method and and 
applies it to the magnetized gas. We concentrate initially on the specific heat and show how the presence
of the magnetic field alters the behavior from that found for the free gas. We study numerically what happens
for large and small magnetic fields. Results are also obtained for the magnetization. In Section \ref{sec:meb}
we describe how it is possible to obtain analytical results for the critical temperature (when $D\geq 5$),
the magnetization, and the specific heat when the magnetic field is weak. We are able to analytically 
confirm the numerical results concerning the limit $B\rightarrow 0$. For the $D=3$ gas we give 
approximations valid at the critical temperature for the free Bose gas. (Previous approximations were 
only valid for temperatures larger than this.) A number of appendices contain details of the expansions used
for the analytical results.

\section{Bose--Einstein condensation with no external field}
\label{sec:becnob}

In this section we wish to review very briefly some of the basic properties of the free Bose gas. 
The spatial dimension $D$ will be arbitrary. For $D=3$ the analysis is standard textbook material 
\cite{lond50a,huan63a,landlifs80a,path72a}. The absence of Bose--Einstein condensation when $D=2$ 
is also widely known \cite{osbo49a,may59a}. Spatial dimensions $D>3$ have also been studied 
\cite{may64a,ziffuhlekac77a}.

We will consider an ideal gas of $N$ spinless bosons confined in a large box of volume $V$ in $D$ 
spatial dimensions. The infinite volume limit will be taken with ${\cal N}=\frac{N}{V}$ fixed as is 
conventionally done. The energy levels of the system are
\begin{equation}
\label{21010}
E_{n_{i}}=\frac{1}{2 m}\sum_{i=1}^{D}\left(\frac{2\pi n_{i}}{L_{i}}\right)^{2}
\end{equation}
before the infinite volume limit is taken if we impose periodic boundary conditions on the field. 
We set $\hbar=1$ throughout the paper. In (\ref{21010}), $n_{i}=0,\pm 1,\pm2,\dots$ and $L_{i}$ is 
the length of the box in the $i$th direction. In section \ref{sec:becwib} we will show how the effective 
action formalism can be used to study this problem (see references \cite{toms94a,kirstoms96b,toms97a} 
for reviews). For now we stick with the conventional thermodynamic expressions. 

The internal energy is given by
\begin{equation}
\label{21020}
U = \sum_{n_{i}} \frac{E_{n_{i}}}{
\left(e^{\beta \left(E_{n_{i}}-\mu\right)}-1\right)}
\end{equation}
where $\mu$ is the chemical potential and $\beta=T^{-1}$ in units with the Boltzmann constant 
$k_{\sss B}=1$. The grand canonical ensemble is used here.

By taking the infinite volume limit, we can replace the sums in (\ref{21020}) over $n_{i}$ with integrals. 
After  expanding the denominator of (\ref{21020}) in a geometric series, and using (\ref{21010}) for the 
energy levels, the integrals over $n_{i}$ may be performed with the result
\begin{equation}
\label{21030}
U = \frac{D}{2}\beta V \left(\frac{m}{2 \pi \beta} \right)^{\frac{D}{2}}\li{\frac{D+2}{2}}{e^{\beta \mu}}.
\end{equation}
We have defined the polylogarithm function $\li{p}{z}$ by
\begin{equation}
\label{21040}
\li{p}{z} = \sum_{k=1}^{\infty}\frac{z^{k}}{k^{p}}.
\end{equation}
We note the property (assuming $p>1$)
\begin{equation}
\label{21050}
\li{p}{1}=\zer{p},
\end{equation}
where $\zer{p}$ denotes the Riemann $\zeta$--function.

The essential feature of Bose--Einstein condensation as a phase transition is exhibited in the behavior of 
the specific heat. The specific heat at constant volume $C_{v}$ is defined by
\begin{equation}
\label{22010}
C_{v} = \left(\frac{\partial U}{\partial T}\right)_{V,N}
= -\beta^{2}\left(\frac{\partial U}{\partial \beta}\right)_{V,N}.
\end{equation}
The particle number $N$, given by
\begin{equation}
\label{22020}
N = V \left(\frac{m}{2 \pi \beta} \right)^{\frac{D}{2}}\li{\frac{D}{2}}{e^{\beta \mu}},
\end{equation}
is held fixed when computing the derivative in (\ref{22010}). The chemical potential is not fixed. From 
(\ref{21030}), (\ref{22010}) we find
\begin{eqnarray}
\nn
C_{v} &=& V \left(\frac{m}{2 \pi\beta}\right)^{\frac{D}{2}}\left[\frac{D(D+2)}{4}
\li{\frac{D+2}{2}}{e^{\beta \mu}}\right. \\
\label{22030}
&&\left. -\frac{D}{2}\beta(\beta\mu)' \li{\frac{D}{2}}{e^{\beta \mu}}\right].
\end{eqnarray}
Here $'$ denotes the derivative with respect to $\beta$ holding $V$ and $N$ fixed. 
To calculate $(\beta\mu)'$ we differentiate both sides of (\ref{22020}) to give 
\begin{equation}
\label{22040}
(\beta\mu)' = \frac{D}{2\beta}\frac{\li{\frac{D}{2}}{e^{\beta \mu}}}{\li{\frac{D-2}{2}}{e^{\beta \mu}}}.
\end{equation}
Substitution of (\ref{22040}) into (\ref{22030}) and use of (\ref{22020}) shows that
\begin{equation}
\label{22050}
\frac{C_{v}}{N} =\frac{D(D+2)}{4} \frac{\li{\frac{D+2}{2}}{e^{\beta \mu}}}
{\li{\frac{D}{2}}{e^{\beta \mu}}}-\frac{D^{2}}{4} 
\frac{\li{\frac{D}{2}}{e^{\beta \mu}}}{\li{\frac{D-2}{2}}{e^{\beta \mu}}}.
\end{equation}

If we confine ourselves to $D\geq 3$, as we do for the rest of the paper, then a critical 
temperature $T_{0}$ can exist at which the chemical potential $\mu=0$. From (\ref{22020}) we find
\begin{equation}
\label{22060}
T_{0} = \left(\frac{2 \pi}{m}\right)\left[\frac{ N}
{V \zeta_{\sss R}\left(\frac{D}{2}\right)}\right]^{\frac{2}{D}}.
\end{equation}
For $T<T_{0}$ the chemical potential remains fixed at $\mu=0$. It is easy to see that (\ref{22050})
only holds for $T>T_{0}$. When $T<T_{0}$, the specific heat may be evaluated from (\ref{22030}) by 
setting the term with $(\beta\mu)'$ to zero and also $\mu$ to zero. This results in
\begin{equation}
\label{22070}
\frac{C_{v}}{N} = \left(\frac{\beta_{0}}{\beta}\right)^{\frac{D}{2}}\frac{D(D+2)}{4} 
\frac{\zer{\frac{D+2}{2}}}{\zer{\frac{D}{2}}}
\end{equation}
for $T<T_{0}$.

The two expressions (\ref{22050}),(\ref{22070}) may be used to compute the specific heat for all 
temperatures, and also to study the behavior at $T=T_{0}$. It is easy to see from the polylogarithm 
function (\ref{21040}) that if $p\leq 1$, $\li{p}{z}\rightarrow\infty$ as $z\rightarrow 1$. This means 
that the second term of (\ref{22050}) vanishes as $\mu\rightarrow 0$ for $D=3,4$, but is finite for 
$D\geq 5$. Thus the specific heat is continuous at $T=T_{0}$ for $D=3,4$, but discontinuous at $T=T_{0}$ 
for $D\geq 5$. The discontinuity for $D\geq 5$ is easily computed in terms of Riemann $\zeta$--functions. 
The continuity of $C_{v}$ for $D=3$ is well known \cite{lond50a,huan63a,landlifs80a,path72a}. 
The behavior for $D>3$ can be found in \cite{may64a,ziffuhlekac77a}.

\section{Bose--Einstein condensation in a constant external magnetic field}
\label{sec:becwib}

When a magnetic field is applied to a gas of charged bosons in three spatial dimensions the energy 
spectrum (in the infinite volume limit) contains a discrete harmonic oscillator--like part as well as a 
continuous part. The discrete part is just the Landau level quantization \cite{landlifs65a}. For $D>3$ 
there may be a number of discrete components because the magnetic field is not described by a vector, 
but by an antisymmetric tensor with more than one independent component \cite{toms95a}. For simplicity 
we will restrict our attention to the case of only a single nonzero component in the present paper. We wish 
to provide a treatment similar to that for the free Bose gas when a nonzero magnetic field is present. In 
particular we will study the specific heat and see how the presence of a magnetic field alters the behavior 
from that found for the free Bose gas in section \ref{sec:becnob}. Also we will compute the magnetization 
and study the Meissner--Ochsenfeld effect in detail. The formalism used is the effective action method as 
reviewed in references \cite{kirstoms96b,toms97a}. This formalism allows the nonzero condensate (if 
there is one) to be treated in a very natural manner.

\subsection{Thermodynamic potential and phase transitions: General formalism}
\label{subsec:genform}

The thermodynamic potential is usually defined by
\begin{equation}
\label{31010}
\Omega_{T\neq 0} = \frac{1}{\beta}\sum_{n}\ln{\left[1-e^{\beta(E_{n}-e\mu)}\right]}.
\end{equation}
However, in the effective action method there is another term present if there is a nonzero condensate 
described by a background field $\bar{\Psi}$. This is 
\begin{equation}
\label{31020}
\Omega^{(0)} = \int d^{\sss D}x \left\{\frac{1}{2m}|\vc{D}\bar{\Psi}|^{2}-e\mu|\bar{\Psi}|^{2}\right\}.
\end{equation}
Here $\vc{D}\bar{\Psi}=\vc{\nabla}\bar{\Psi}-ie\vc{A}\bar{\Psi}$ is the usual gauge--covariant derivative. 
The complete thermodynamic potential is
\begin{equation}
\label{31030}
\Omega=\Omega^{(0)}+\Omega_{T\neq 0}.
\end{equation}
(Actually, if we are interested in the dynamics of the magnetic field there will be an additional term 
involving a Maxwell action. We will consider this in section \ref{subsec:mag} below). Given the 
thermodynamic potential, all quantities of interest can be calculated.

The presence of a condensate is signalled by a nonzero value for $\bar{\Psi}$. This is associated with 
symmetry breaking as discussed in the relativistic case \cite{kapu81a,habeweld82a}. In our case 
$\bar{\Psi}$  must satisfy
\begin{equation}
\label{31040}
\frac{\delta \Omega}{\delta \bar{\Psi}}=0=\frac{1}{2m}\vc{D}^{2}\bar{\Psi}+e\mu\bar{\Psi}.
\end{equation}
We can solve this by expanding $\bar{\Psi}(\vc{x})$ in terms of the stationary state solutions to the 
Schr\"{o}dinger equation:
\begin{equation}
\label{31050}
\frac{-1}{2m}\vc{D}^{2}f_{n}(\vc{x})=E_{n}f_{n}(\vc{x}).
\end{equation}
If we write
\begin{equation}
\label{31060}
\bar{\Psi}(\vc{x})=\sum_{n}C_{n}f_{n}(\vc{x})
\end{equation}
for some coefficients $C_{n}$, and assume that the set of solutions $f_{n}(\vc{x})$ forms a complete set, 
then (\ref{31040}) results in
\begin{equation}
\label{31070}
0=(E_{n}-e\mu)C_{n}.
\end{equation}
We will define a critical value of $\mu$, say $\mu_{\sss C}$, by
\begin{equation}
\label{31080}
e\mu_{\sss C}=E_{0},
\end{equation}
where $E_{0}$ is the lowest energy level. If $\mu<\mu_{\sss C}$, then the only solution to (\ref{31070}) is 
for $C_{n}=0$, which corresponds to $\bar{\Psi}=0$. There is no condensate in this case associated with 
symmetry breaking and a phase transition. However if $\mu$ can reach the value $\mu_{\sss C}$ defined 
in (\ref{31080}) for some temperature $T_{\sss C}$, then $C_{0}$ in (\ref{31070}) is undetermined and 
we can have a nonzero condensate described by
\begin{equation}
\label{31090}
\bar{\Psi}(\vc{x})=C_{0}f_{0}(\vc{x}).
\end{equation}
The temperature $T_{\sss C}$ at which $\mu=\mu_{\sss C}$ is called the critical temperature. 

For the case of the free gas considered in section \ref{sec:becnob} we have $E_{0}=0$, so that 
$\mu_{\sss C}=0$. The critical temperature $T_{\sss C}$ is then the value of the temperature at which 
the chemical potential vanishes as stated earlier. If the spatial dimension $D\geq 3$ a critical temperature 
exists and signals a phase transition with a nonzero value for $\bar{\Psi}$. ($\bar{\Psi}$ is constant for 
the free Bose gas). Associated with this phase transition is a growth in the number of particles in the 
ground state.

For some systems it is possible to have a sudden growth in the occupancy of the ground state without a 
phase transition. In this case $\mu$ never reaches the critical value of $\mu_{\sss C}$ but instead 
approaches it asymptotically. The speed at which the ground state particle number builds up depends on 
how fast $\mu$ approaches $\mu_{\sss C}$. Because $\mu$ never reaches $\mu_{\sss C}$ we have 
$\bar{\Psi}=0$, and no symmetry breaking. As we will see below, for a charged Bose gas in a constant 
magnetic field the spatial dimension $D$ determines whether or not a phase transition occurs. 

We can use our expression (\ref{31030}) for $\Omega$ to find the total charge $Q$, since 
$Q=-\frac{\partial \Omega}{\partial \mu}$ with $V, B, \beta,$ and $\bar{\Psi}$ held fixed. It is convenient 
to write
\begin{equation} 
\label{31100}
Q= Q_{0} + Q_{1}
\end{equation}
where
\begin{equation} 
\label{31110}
Q_{0}= -\frac{\partial \Omega^{(0)}}{\partial \mu}=e\int d^{\sss D}|\bar{\Psi}|^{2}=e|C_{0}|^{2},
\end{equation}
if we use (\ref{31020}), (\ref{31090}), and
\begin{equation} 
\label{31120}
Q_{1}= -\frac{\partial \Omega_{T\neq 0}}{\partial \mu}.
\end{equation}
From (\ref{31010}) we find
\begin{equation} 
\label{31130}
Q_{1}= e\sum_{n}\frac{1}{\left[e^{\beta(E_{n}-e\mu)}-1\right]}.
\end{equation}
If we can always solve $Q=Q_{1}$ for $\mu$ for all temperatures, then $\bar{\Psi}=0$. There is no 
condensate, symmetry breaking, or phase transition in this case. If it is not possible to solve $Q=Q_{1}$ 
for $\mu$, then we must have $Q_{0}\neq 0$ and find a nonzero value for $\bar{\Psi}$.

\subsection{Thermodynamic potential: Constant magnetic field}
\label{subsec:potwib}

The formalism outlined in section \ref{subsec:genform} will now be applied to the $D$--dimensional 
charged Bose gas in a constant one--component magnetic field. We will assume $D\geq 3$ here and pick the 
magnetic field in the $z$--direction. It is possible to solve (\ref{31050}) for the energy levels and 
corresponding eigenfunctions \cite{landlifs65a}. We have (choosing $\hbar=c=1$)
\begin{equation}
\label{32010}
E_{n,k_{i}}=\left(n+\frac{1}{2}\right)\omega
+\frac{1}{2m}\sum_{i=3}^{D}\left(\frac{2\pi k_{i}}{L_{i}}\right)^{2}
\end{equation}
where $n=0,1,\dots$ labels the Landau level, and $k_{i}=0,\pm1,\dots$ if we impose periodic boundary 
conditions on a box as in section \ref{sec:becnob}. We have defined
\begin{equation}
\label{32020}
\omega=\frac{eB}{m}.
\end{equation}
The energy levels (\ref{32010}) are degenerate with degeneracy
\begin{equation}
\label{32030}
g=\frac{eBL_{1}L_{2}}{2\pi}.
\end{equation}

As in section \ref{sec:becnob} we will be interested in the large box limit with $L_{i}\rightarrow\infty$. 
In this limit we can replace the sums over $k_{i}$ resulting when (\ref{32010}) is used in (\ref{31010})
with integrals. A change of variables gives
\begin{eqnarray}
\nn
\Omega_{T\neq 0} &=& \frac{eBV}{2\pi\beta}\sum_{n=0}^{\infty}\int\frac{d^{{\sss D}-2} k}{(2\pi)^{{\sss D}-2}} \\
\label{32040}
&& \times \ln{\left\{1-e^{-\beta\left[\left(n+\frac{1}{2}\right)\omega+\frac{k^{2}}{2m}-e\mu\right]}\right\}}.
\end{eqnarray}
This may be evaluated by expanding the logarithm in its Taylor series and then performing the integral 
over $k$. We find
\begin{eqnarray}
\nn
\Omega_{T\neq 0} &=& -\omega V \left(\frac{m}{2 \pi \beta} \right)^{\frac{D}{2}}
\sum_{n=0}^{\infty} \sum_{l=1}^{\infty}l^{-\frac{D}{2}} 
e^{-l\beta\left[\left(n+\frac{1}{2}\right)\omega-e\mu\right]} \\
\label{32060}
&=&-\omega V \left(\frac{m}{2 \pi \beta} \right)^{\frac{D}{2}}
\sum_{l=1}^{\infty}\frac{l^{-\frac{D}{2}} e^{-l\beta\left(\frac{\omega}{2}-e\mu\right)}}
{(1-e^{-l\beta\omega})}
\end{eqnarray}
where the second line has followed by performing the sum over $n$.

At this stage it is useful to define a dimensionless temperature. We define
\begin{equation}
\label{32070}
x=\beta\omega.
\end{equation}
$x$ is seen to be the ratio between the energy gap between successive energy levels $\omega$ and the 
thermal energy $k_{\sss B} T$. The lowest energy level from (\ref{32010}) is $E_{0,0}=\frac{\omega}{2}$. 
From (\ref{31080}) the critical value for $\mu$ is
\begin{equation}
\label{32080}
e\mu_{\sss C}=\frac{\omega}{2}.
\end{equation}
We will define a dimensionless chemical potential $\varepsilon$ by
\begin{equation}
\label{32090}
e\mu=\omega\left(\frac{1}{2}-\varepsilon\right).
\end{equation}
A phase transition is characterized by a critical temperature $T_{\sss C}$ at which $\varepsilon=0$. 
The expression (\ref{32060}) may be written in terms of the dimensionless variables $x$ and $\varepsilon$:
\begin{equation}
\label{32100}
\Omega_{T\neq 0} = -\omega V \left(\frac{m}{2 \pi \beta} \right)^{\frac{D}{2}}
\sum_{l=1}^{\infty}\frac{l^{-\frac{D}{2}} e^{-l\varepsilon x}}{(1-e^{-lx})}.
\end{equation}
It is convenient to introduce some compact notation for the class of sums we will encounter
in order to simplify formulae. Let
\begin{equation}
\label{32110}
\bes{\kappa}{\alpha}{\delta} = \sum_{l=1}^{\infty} \frac{l^{\frac{\alpha}{2}} e^{-l x (\varepsilon+\delta)}}
{\left(1-e^{-l x}\right)^{\kappa}}.
\end{equation}
With this notation we may write (\ref{32100}) as
\begin{equation}
\label{32120}
\Omega_{T\neq 0} = -\omega V \left(\frac{m}{2 \pi \beta} \right)^{\frac{D}{2}}\bes{1}{-D}{0}.
\end{equation}
Various thermodynamic quantities involve derivatives of the thermodynamic potential. We will initially 
consider the charge.

From (\ref{31120}), using (\ref{32120}) we have
\begin{eqnarray}
\nn
Q_{1} &=& \omega V \left(\frac{m}{2 \pi \beta} \right)^{\frac{D}{2}}
\frac{\partial}{\partial \mu} \bes{1}{-D}{0}\\
\label{32130}
&=& e V \left(\frac{m}{2 \pi \beta} \right)^{\frac{D}{2}}
x \bes{1}{2-D}{0}.
\end{eqnarray}
Whether or not a phase transition occurs is determined by the convergence or divergence of $\bes{1}{2-D}{0}$
as $\varepsilon\rightarrow 0$. If this sum diverges as $\varepsilon\rightarrow 0$ then we will always be 
able to solve $Q=Q_{1}$ for $\mu$. As discussed in section \ref{subsec:genform} this means that there is 
no phase transition. From (\ref{32110}) it is easy to see that $\bes{1}{\alpha}{0}$ diverges as 
$\varepsilon\rightarrow 0$ for $\alpha\geq -2$. With $\alpha=2-D$, this means that there is no phase 
transition for $D=3, 4$ dimensions. For $D\geq 5$, $\bes{1}{2-D}{0}$ converges as 
$\varepsilon\rightarrow 0$ and there is a phase transition. In this case ({\em i.e.} for $D\geq 5$) there is a 
nonzero condensate characterized by $\bar{\Psi}\neq 0$.

For $D\geq 5$ we may use (\ref{31090}) to find $\bar{\Psi}$. If we choose the gauge
\begin{equation}
\label{32140}
A_{1}=-B y, \;\;\;\; A_{2}=\dots=A_{\sss D}=0,
\end{equation}
for the vector potential, then
\begin{equation}
\label{32150}
f_{0}=\alpha e^{-\frac{1}{2}eBy^{2}}
\end{equation}
is the eigenfunction corresponding to the lowest energy $E_{0}=\frac{\omega}{2}$. $\alpha$ is a constant 
chosen to normalize $f_{0}$. $\Omega^{(0)}$ is found by using (\ref{31090}) with (\ref{32150}) in the general 
expression (\ref{31020}). We will return to this when we discuss the magnetization in section \ref{subsec:mag}.

\subsection{The specific heat}
\label{subsec:cv}

The internal energy is given by $U=\frac{\partial}{\partial \beta}(\beta\Omega)$ with 
$V,\omega,$ and $\beta\mu$ held fixed. Using (\ref{31010}---\ref{31030}) we find
\begin{equation}
\label{33010}
U=\int d^{\sss D}x\left\{\frac{1}{2m}|\vc{D}\bar{\Psi}|^{2}\right\}
+\frac{\partial}{\partial \beta}\left(\beta\Omega_{T\neq 0}\right).
\end{equation}
The first term accounts for any nonzero condensate, and the second term is easily seen to be
\begin{equation}
\label{33020}
\frac{\partial \left(\beta\Omega_{T\neq 0}\right)}{\partial \beta}
=\sum_{n}\frac{E_{n}}{\left[e^{\beta(E_{n}-e\mu)}-1\right]}
\end{equation}
which is the usual expression for the internal energy (see equation (\ref{21020}) with a different definition 
for $\mu$). With (\ref{31050}) and (\ref{31090}) we find
\begin{equation}
\label{33030}
U=E_{0}|C_{0}|^{2}+\frac{\partial}{\partial \beta}\left(\beta\Omega_{T\neq 0}\right).
\end{equation}
If $E_{0}=0$, then the contribution from $\Omega^{(0)}$ to the internal energy vanishes, and the internal energy 
is given by the standard expression (\ref{33020}). This is the situation for the free Bose gas discussed in 
section \ref{sec:becnob}. For the constant magnetic field, $E_{0}=\frac{\omega}{2}$, so that if 
$\bar{\Psi}\neq 0$ we must include the condensate contribution to obtain the correct expression for the 
energy.

If we use (\ref{31110}), then (\ref{33030}) may be written as 
\begin{equation}
\label{33040}
U=\frac{\omega Q_{0}}{2e}+\frac{\partial}{\partial \beta}\left(\beta\Omega_{T\neq 0}\right).
\end{equation} 
With (\ref{32120}) for $\Omega_{T\neq 0}$, and differentiating with respect to $\beta$ keeping 
$\beta\mu$, $\omega$, $V$ fixed results in
\begin{eqnarray}
\nn
\frac{\partial \left(\beta\Omega_{T\neq 0}\right)}{\partial \beta}
&=& \omega V \left(\frac{m}{2 \pi \beta} \right)^{\frac{D}{2}}\left\{\rule{0mm}{5mm}x\bes{2}{2-D}{1} \right. \\
\nn
&&+ \frac{x\bes{1}{2-D}{0}}{2} \\
\label{33050}
&&\left. + \frac{(D-2)\bes{1}{-D}{0}}{2}  \right\}.
\end{eqnarray}
Noting (\ref{32130}) allows us to write this as 
\begin{eqnarray}
\nn
\frac{\partial}{\partial \beta}\left(\beta\Omega_{T\neq 0}\right) &=& 
\frac{\omega Q_{1}}{2e}+\omega V \left(\frac{m}{2 \pi \beta} \right)^{\frac{D}{2}}
\left\{\rule{0mm}{5mm}x \bes{2}{2-D}{1} \right. \\
\label{33060}
&&\left. +\left(\frac{(D-2)}{2}\right)\bes{1}{-D}{0}\right\}.
\end{eqnarray}
Substitution of (\ref{33060}) back into (\ref{33040}) and noting from (\ref{31100}) that $Q=Q_{0}+Q_{1}$ 
where $Q$ is the total charge, results in
\begin{eqnarray}
\nn
U &=& \frac{\omega Q}{2e}+\omega V \left(\frac{m}{2 \pi \beta} \right)^{\frac{D}{2}}
\left\{\rule{0mm}{5mm}x \bes{2}{2-D}{1} \right. \\
\label{33070}
&&\left. +\left(\frac{(D-2)}{2}\right)\bes{1}{-D}{0}\right\}.
\end{eqnarray}
The expression for the internal energy we have just obtained (\ref{33070}) holds whether there is a nonzero 
condensate ($\bar{\Psi}\neq 0$) or not. In cases where a phase transition does occur, the r\^{o}le of the 
condensate is crucial for obtaining the correct expression for the internal energy. If we had neglected the 
contribution coming from $\Omega^{(0)}$ then (\ref{33070}) would have $Q_{1}$ in place of $Q$. This would 
then lead to an erroneous expression for the specific heat, since $Q$ is fixed whereas $Q_{1}$ is not.

\begin{figure}
\setlength{\epsfxsize}{80mm}
\centerline{\epsfbox{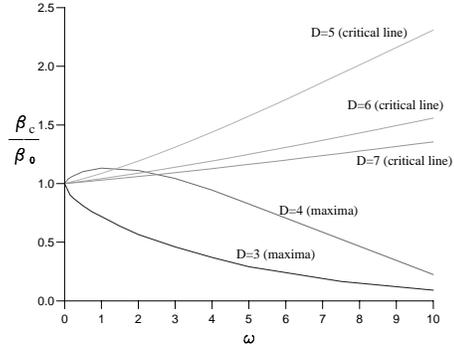}}
\caption{For $D\geq 5$ this shows the value of $\beta_{\sss C}$ compared to the free Bose gas critical 
value of $\beta_{0}$. For $D=3, 4$ because there is no critical temperature when a magnetic field is 
present we have plotted the inverse temperature of the specific heat maximum in units of $\beta_{0}$.}
\label{figcrit}
\end{figure}

The specific heat at constant volume was defined in (\ref{22010}). The quantities held fixed are 
$V, Q, \bar{\Psi},$ and $\omega=\frac{eB}{m}$ when the differentiation is performed. This means that the 
first term in (\ref{33070}) makes no contribution to $C_{v}$. Just like the free Bose gas discussed in 
section \ref{sec:becnob}, we must distinguish between the expressions above and below the the critical 
temperature if there is a phase transition. For $D\geq 5$, the critical temperature $T_{\sss C}$ is 
defined by
\begin{equation}
\label{33080}
Q= eV \left.\left(\frac{m}{2 \pi \beta_{\sss C}} \right)^{\frac{D}{2}} x_{\sss C} \bes{1}{2-D}{0}
\right|_{\sstack{\varepsilon=0,}{x=x_{\sss C}}}.
\end{equation}
Here $x_{\sss C}= \beta_{\sss C}\omega$ with $\beta_{\sss C}= T_{\sss C}^{-1}$, and the sum 
$\bes{1}{2-D}{0}$ is evaluated with $\varepsilon=0$ and $x=x_{\sss C}$. Unlike for the free Bose gas, 
it is not possible to solve for $T_{\sss C}$ analytically. We will return to an approximate evaluation of 
$T_{\sss C}$ in section \ref{sec:meb}. We have solved (\ref{33080}) numerically to find $T_{\sss C}$. The 
results for small values of $\omega$ are shown in figure \ref{figcrit}. As expected, when 
$\omega\rightarrow 0$ we have $T_{\sss C}\rightarrow T_{0}$.

For small values of $\omega$, $\beta_{\sss C}$ is close to $\beta_{0}$ (an approximate analytical 
expression will be given in section \ref{sec:meb}). As the strength of the magnetic field is increased the 
deviation of $T_{\sss C}$ from $T_{0}$ becomes more pronounced, but lessens with increasing dimension. 
In all cases where a phase transition occurs, the critical temperature is lower than the result for the free 
Bose gas as observed in \cite{toms95b}.

Let $C_{v}^{>}$ denote the specific heat for $T>T_{\sss C}$ and $C_{v}^{<}$ be the specific heat for 
$T<T_{\sss C}$. Using (\ref{33070}) we have
\begin{eqnarray}
\nn
C_{v}^{>} &=& x V \left(\frac{m}{\rule{0mm}{5mm}2 \pi \beta}\right)^{\frac{D}{2}}\left\{2 x^{2} \bes{3}{4-D}{2} \right. \\
\nn
&& + x^{2} \bes{2}{4-D}{1} + (D-2) x \bes{2}{2-D}{1} \\
\nn
&& + \frac{D(D-2)}{4} \bes{1}{-D}{0} + \beta x \left(\frac{\partial(\varepsilon x)}{\partial \beta}\right)_{Q} \\
\label{33090}
&& \left. \times \left[\bes{2}{4-D}{1} +\left(\frac{D-2}{2}\right) \bes{1}{2-D}{0}\right] \right\}.
\end{eqnarray}
The terms in $\left(\frac{\partial(\varepsilon x)}{\partial \beta}\right)_{Q}$ arise from the fact that the 
charge $Q$ rather than the chemical potential $\mu$ is held fixed. By setting $Q=Q_{1}$ in (\ref{32130})
(valid since $T>T_{\sss C}$), and differentiating both sides with respect to $\beta$ with $Q$ held fixed 
gives
\begin{eqnarray}
\nn
\beta x \left(\frac{\partial(\varepsilon x)}{\partial \beta}\right)_{Q} &=& -\frac{x}{\bes{1}{4-D}{0}}
\left(\rule{0mm}{5mm}x \bes{2}{4-D}{1} \right. \\
\label{33100}
&&\left. +\left(\frac{D-2}{2}\right) \bes{1}{2-D}{0} \right).
\end{eqnarray}
Substitution of (\ref{33100}) into (\ref{33090}) leads to
\begin{eqnarray}
\nn
C_{v}^{>} &=& x V \left(\frac{m}{2 \pi \beta}\right)^{\frac{D}{2}}
\left\{\rule{0mm}{5mm}2 x^{2} \bes{3}{4-D}{2} + x^{2} \bes{2}{4-D}{1} \right. \\
\nn
&&  + (D-2) x \bes{2}{2-D}{1}  + \frac{D(D-2)}{4} \bes{1}{-D}{0} \\
\nn
&& - \frac{x (D-2) \bes{2}{4-D}{1} \bes{1}{2-D}{0}}{\protect{\bes{1}{4-D}{0}}} \\
\nn
&& -\frac{x^{2} (\bes{2}{4-D}{1})^{2}}{\protect{\bes{1}{4-D}{0}}} \\
\label{33110}
&&\left. -\left(\frac{D-2}{2}\right)^{2}\frac{(\bes{1}{2-D}{0})^{2}}{\protect{\bes{1}{4-D}{0}}} \right\}.
\end{eqnarray}
For $T<T_{\sss C}$ we have $\varepsilon=0$ fixed. (Equivalently, $\mu=\frac{\omega}{2}$ is fixed). 
The result of this is that $C_{v}^{<}$ is given by (\ref{33090}) but with the last two terms set to zero, 
and $\varepsilon=0$ in all other terms:
\begin{eqnarray}
\nn
C_{v}^{<} &=& x V \left(\frac{m}{2 \pi \beta}\right)^{\frac{D}{2}}\left\{\rule{0mm}{5mm}2 x^{2} \bes{3}{4-D}{2} \right. \\
\nn
&& + x^{2} \bes{2}{4-D}{1} + (D-2) x \bes{2}{2-D}{1} \\
\label{33120}
&&\left. + \frac{D(D-2)}{4} x \bes{1}{-D}{0} \right\}\rule[-4mm]{0.1mm}{9mm}_{\;\varepsilon=0}.
\end{eqnarray}
By comparing $C_{v}^{>}$ in (\ref{33110}) with $C_{v}^{<}$ in (\ref{33120}) it can be seen that whether or not 
the specific heat is continuous at the critical temperature is determined by the behavior of $\bes{1}{4-D}{0}$ 
as $\varepsilon\rightarrow 0$. The two expressions will only agree if $\bes{1}{4-D}{0}$ diverges in this 
limit. From the definition (\ref{32110}) this only happens for $D=5, 6$ (Recall that we are assuming 
$D\geq 5$ here so that $T_{\sss C}$ exists). We conclude that the specific heat for the charged Bose gas 
in a constant magnetic field is continuous at the critical temperature for $D=5, 6$ and discontinuous for 
$D\geq 7$.

In the cases $D=3, 4$ where there is no phase transition, $\varepsilon$ never vanishes. The specific heat 
is given by (\ref{33110}) in these two cases for all temperatures. When $D=3, 4$ the specific heat is a 
perfectly smooth function of temperature. 

\begin{figure}
\setlength{\epsfxsize}{80mm}
\centerline{\epsfbox{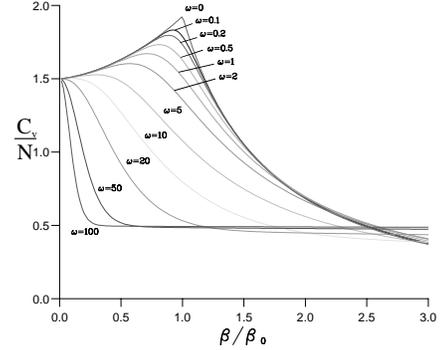}}
\caption{This shows the specific heat at constant volume in units of the total number of particles as 
a function of the inverse temperature in units of $\beta_{0}$ where $\beta_{0}=T_{0}^{-1}$ with $T_{0}$ 
given in (\ref{22060}) for $D=3$. For comparison the free gas result is also shown ($\omega=0$).}
\label{fig2}
\end{figure}

\begin{figure}
\setlength{\epsfxsize}{80mm}
\centerline{\epsfbox{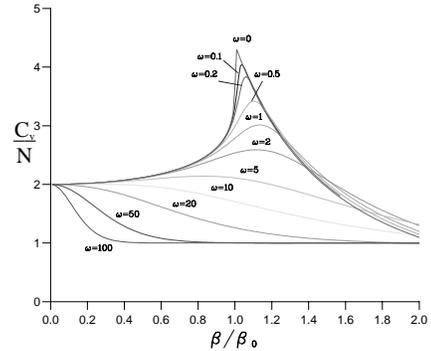}}
\caption{The specific heat at constant volume in units of the total number of particles for the magnetized 
gas with $D=4$. The top curve ($\omega=0$) is the result for the free Bose gas.}
\label{fig3}
\end{figure}

\begin{figure}
\setlength{\epsfxsize}{80mm}
\centerline{\epsfbox{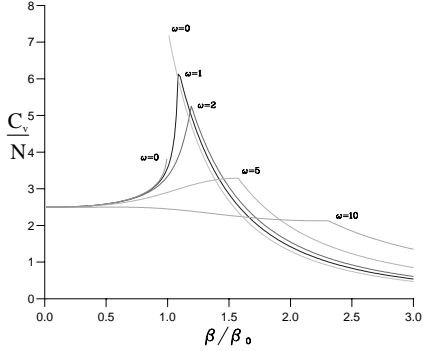}}
\caption{The specific heat at constant volume in units of the total number of particles for the 
magnetized gas with $D=5$. The curves are given as a function of $\beta=T^{-1}$ in units of $\beta_{0}$ 
rather than $\beta_{\sss C}$. The discontinuous free gas result is labeled $\omega=0$.}
\label{fig4}
\end{figure}

\begin{figure}
\setlength{\epsfxsize}{80mm}
\centerline{\epsfbox{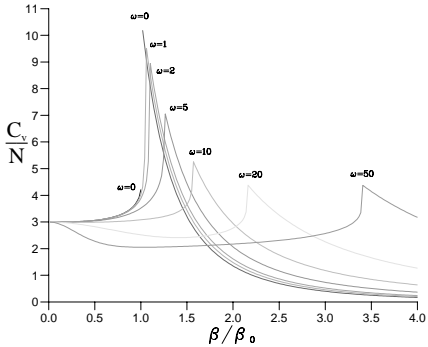}}
\caption{The specific heat at constant volume in units of the total number of particles for the magnetized 
gas with $D=6$. The discontinuous free gas result is shown as $\omega=0$.}
\label{fig5}
\end{figure}

\begin{figure}
\setlength{\epsfxsize}{80mm}
\centerline{\epsfbox{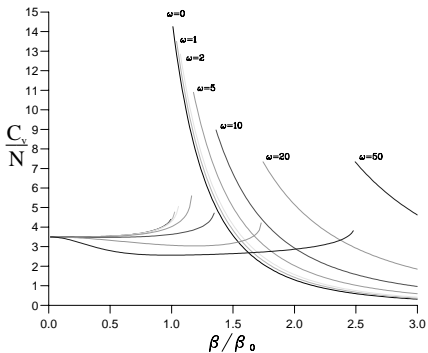}}
\caption{The specific heat at constant volume in units of the total number of particles for the magnetized 
gas with $D=7$. This is the first spatial dimension for which the specific heat has a discontinuity at the 
critical temperature. The free gas result ($\omega=0$) is also shown .}
\label{fig6}
\end{figure}

Graphs showing the specific heat for $D=3\mbox{---}7$ are shown in figures \ref{fig2}---\ref{fig6}.

For the $D=3, 4$ gases, the presence of the magnetic field is seen in figures \ref{fig2},\ref{fig3} to round 
off the familiar sharp behavior at the transition temperature for the free gas. In both cases as the magnetic 
field is reduced the curves for the specific heat tend towards the free--field results. The maximum in the 
specific heat becomes sharper, and the temperature at which the maximum occurs tends towards the 
value $T_{0}$ as the magnetic field is reduced. For $B\neq 0$, although there is no phase transition 
characterized by a critical temperature and a nonzero condensate, as $B$ is reduced the specific heat 
starts to look more and more like the free gas result. We will look at this analytically in section 
\ref{sec:meb}. In both cases, as $\beta\rightarrow 0$ (or $T\rightarrow \infty$) the specific heats for 
$D=3, 4$ approach the classical Maxwell--Boltzmann results of 1.5 (for $D=3$) and 2 (for $D=4$).

The behavior of the specific heat as $B$ is increased is also of interest. Increasing the value of $B$ tends 
to reduce the specific heat maximum and to broaden the curves. In fact for large values of $B$ if we 
examine figure \ref{fig2} it can be seen that the curves approach the value of $\frac{1}{2}$ as $B$ is 
reduced before rising sharply to the classical result of $\frac{3}{2}$. This demonstrates that the specific
heat for the gas with $D=3$ resembles the specific heat for the free gas with $D=1$ in a strong magnetic 
field. The value of $\frac{1}{2}$ is the classical value 
for the one--dimensional gas. For the magnetized gas with $D=4$, figure \ref{fig2} shows that the specific 
heat approaches the value of 1 in  a strong magnetic field before the sharp rise to the classical value of 2. 
Again this shows the reduction in the dimension by 2 since $\frac{C_{v}}{N}\rightarrow 1$ as 
$\beta\rightarrow 0$ for the free Bose gas in two spatial dimensions. 
This is totally consistent with the approach used in references \cite{kirstoms96a,kirstoms97a} in 
which the leading behavior of thermodynamic quantities was studied in a general setting by using the 
lowest energy solutions. As $B$ is increased, the gap between the ground state and the excited states 
becomes larger; thus the leading contribution would be expected to come from the ground state.

\begin{figure}
\setlength{\epsfxsize}{80mm}
\centerline{\epsfbox{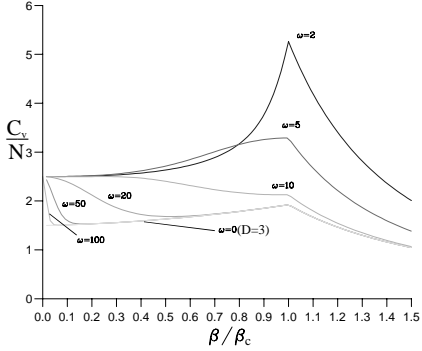}}
\caption{The specific heat is shown in units of $N$ for the magnetized gas with $D=5$ for large values of 
the magnetic field. The curves resemble the free gas specific heat in three spatial dimensions over a wide 
temperature range (compare with $\omega=0$ in figure \ref{fig2}).}
\label{fig7}
\end{figure}

The results for the specific heat of the 5--dimensional gas are shown in figure \ref{fig4}. In this case the 
specific heat for the free Bose gas is discontinuous. Nevertheless as the magnetic field is reduced the 
specific heat curves approach the free gas result. The peaks of the specific heat start to become sharper 
and the slope of the curve steeper. The classical Maxwell--Boltzmann result of $\frac{5}{2}$ is reached 
as $\beta\rightarrow 0$. Just as for the cases $D=3, 4$, the gas exhibits a reduction in the effective 
dimension for large 
values of the magnetic field. This time we would expect to find the specific heat curves looking more and 
more like the familiar form for the specific heat of the free gas in three spatial dimensions. For clarity 
we have given the large magnetic field results on a separate graph in figure \ref{fig7} over a large 
temperature range.

The reduction in dimension should be evident with the specific heat approaching the value $\frac{3}{2}$, the 
classical free gas result for $D=3$, before rising to the value $\frac{5}{2}$ as $\beta\rightarrow 0$.

The specific heats for $D=6, 7$ are shown in figures \ref{fig5},\ref{fig6} respectively. For $D=6$, we see 
the peaks sharpen as the free gas result is approached. Like the 5--dimensional gas the curves are always 
continuous, but the slope becomes greater as $B$ is reduced. For strong magnetic fields the apparent 
reduction in dimension again takes place with the specific heat resembling the free Bose gas in four spatial dimensions. 
The results for $D=7$ are shown in figure \ref{fig6}, and are included because this is the first spatial 
dimension for which the specific heat in the presence of a magnetic field is discontinuous. For strong 
magnetic fields the specific heat resembles the free 5--dimensional gas, and for weak fields the free 
Bose gas limit is approached.

\subsection{Magnetization}
\label{subsec:mag}

Even though for $D=3$ there is no phase transition which can be associated with Bose--Einstein condensation, 
by studying the magnetization of the charged gas Schafroth \cite{scha55a} showed that the gas exhibited the 
Meissner--Ochsenfeld effect. The generalization to other spatial dimensions was performed later 
\cite{may59a,may65a,ariajoan89a,daicfran97a,daicfrangailkowa94a,toms94a,daicfran96a,roja96a,kirstoms97a}. 
We will show how the formalism described in section \ref{subsec:genform} can be used to obtain the 
magnetization. In particular, the r\^{o}le of the condensate $\bar{\Psi}$ for $D\geq 5$ will be examined 
carefully.

The simplest way to see the effects of magnetization is by studying how the field equations for 
electromagnetism are affected. To do this we must include a term in the thermodynamic potential 
(\ref{31030}) for the electromagnetic field. We will use Heaviside--Lorentz rationalized units as usual in 
quantum field theory. (A discussion of the various units and how this alters the expression for the 
magnetization is given in \cite{kirstoms97a}. Of course the physics of the situation should be independent 
of this arbitrary choice). We add on 
\begin{equation}
\label{34010}
\Omega_{em}=\int d^{D}x \left(\frac{1}{4}F_{ij}F^{ij}-J_{ext}^{i}A_{i}\right)
\end{equation}
to (\ref{31030}), where $F_{ij}=\partial_{i}A_{j}-\partial_{j}A_{i}$ is the field strength tensor describing 
the magnetic field, and $J_{ext}^{i}$ is the externally applied current which is responsible for setting up 
the magnetic field. The complete thermodynamic potential is
\begin{equation}
\label{34020}
\Omega = \Omega_{em}+ \Omega^{(0)}+\Omega_{T\neq 0}
\end{equation}
where $\Omega^{(0)}$ and $\Omega_{T\neq 0}$ are given in (\ref{31010}) and (\ref{31020}).

Variation of $\Omega$ with respect to the magnetic field $F_{ij}$ results in 
\begin{equation}
\label{34030}
\partial_{j}H^{ij}=J_{ext}^{i}
\end{equation}
where
\begin{equation}
\label{34040}
H^{ij}=F^{ij} + 2 \frac{\delta}{\delta F_{ij}}\left(\Omega^{(0)}+\Omega_{T\neq 0}\right).
\end{equation}
$H^{ij}$ is the $D$--dimensional analogue of the usual vector $H$ in 3-dimensional electromagnetism. 
As explained earlier it is necessary to treat the magnetic field as a tensor if $D\neq 3$. For more details 
of this analysis see reference \cite{toms97a}.

We have treated the magnetic field generally in (\ref{34010}---\ref{34040}). 
Specializing now to a single component field of strength $B\;\;(F_{12}=-F_{21}=B)$, equation (\ref{34040})
can be written as 
\begin{equation}
\label{34050}
H=B-M,
\end{equation}
where
\begin{equation}
\label{34060}
M=-\frac{\delta}{\delta B}\left(\Omega^{(0)}+\Omega_{T\neq 0}\right),
\end{equation}
and $H^{12}=-H^{21}=H$. Equation (\ref{34050}) is the conventional $B$--$H$ relation found in three spatial 
dimensions, but with the notation defined here it can be seen to hold for all $D$. $M$ in (\ref{34060}) is the 
magnetization. This approach is seen to avoid any ambiguity between what Schafroth called the acting and 
microscopic fields. 

We can now split $M$ in (\ref{34060}) into two pieces in an obvious way. The derivative in (\ref{34060}) is a 
functional derivative, and because $B$ is a constant for our problem we can define
\begin{eqnarray}
\label{34070}
M^{(0)}&=&-\frac{1}{V}\frac{\partial \Omega^{(0)}}{\partial B}, \\
\label{34080}
M_{T\neq 0} &=& -\frac{1}{V}\frac{\partial \Omega_{T\neq 0}}{\partial B}.
\end{eqnarray}
$M_{T\neq 0}$ is easily computed using (\ref{32120}) to be
\begin{eqnarray}
\nn
M_{T\neq 0}&=& \frac{e}{m} \left(\frac{m}{2 \pi \beta}\right)^{\frac{D}{2}} 
\left\{\bes{1}{-D}{0}-\frac{x}{2} \bes{1}{2-D}{0}  \right. \\
\label{34090}
&& \left. - x \bes{2}{2-D}{1} \right\}.
\end{eqnarray}
Recalling (\ref{32130}) we have
\begin{eqnarray}
\nn
M_{T\neq 0}&=&-\frac{1}{2m}\frac{Q_{1}}{V} +\frac{e}{m} \left(\frac{m}{2 \pi \beta}\right)^{\frac{D}{2}}  
\left\{\bes{1}{-D}{0} \right. \\
\label{34100}
&&\left. - x \bes{2}{2-D}{1} \rule{0mm}{5mm}\right\}.
\end{eqnarray}
In cases where no phase transition occurs, we have $Q_{1}=Q$ so that the first term of (\ref{34100}) is 
constant. Also if $\bar{\Psi}=0$ then $M^{(0)}=0$, so that the total magnetization is given by (\ref{34100})
with $Q_{1}=Q$.

When a phase transition does occur we need $M^{(0)}$. With our gauge choice (\ref{32140}), and using 
(\ref{31020}) for $\Omega^{(0)}$, we find
\begin{equation}
\label{34110}
 M^{(0)}= -\frac{e^{2} B}{mV}\int d^{\sss D}x y^{2} |\bar{\Psi}|^{2}.
\end{equation}
$\bar{\Psi}$ was given by (\ref{31090}) with (\ref{32150}). It is easy to show that
\begin{equation}
\label{34120}
 \int d^{\sss D}x y^{2} |f_{0}|^{2}=\frac{1}{2eB},
\end{equation}
which results in
\begin{equation}
\label{34130}
 M^{(0)}= -\frac{e}{2mV}|C_{0}|^{2}=-\frac{Q_{0}}{2mV}
\end{equation}
after using (\ref{31110}). We may now combine (\ref{34100}) and (\ref{34130}) to read
\begin{eqnarray}
\nn
M&=& -\frac{Q}{2mV}+\frac{e}{m} \left(\frac{m}{2 \pi \beta}\right)^{\frac{D}{2}} \left\{\bes{1}{-D}{0} 
\right. \\
\label{34140}
&& \left. - x \bes{2}{2-D}{1} \right\}
\end{eqnarray}
since $Q=Q_{0}+Q_{1}$ is the total charge.

The result in (\ref{34140}) is the exact expression for the magnetization which holds even if there is a phase 
transition. Had the condensate $\bar{\Psi}$ been ignored, we would have obtained (\ref{34100}) rather than 
(\ref{34140}). In the true expression (\ref{34140}) the first term is constant, whereas in (\ref{34100}) the first 
 term is not constant if there is a phase transition. Thus, neglect of the condensate for $D\geq 5$ would lead 
 to an erroneous result for the magnetization.
 
\begin{figure}
\setlength{\epsfxsize}{80mm}
\centerline{\epsfbox{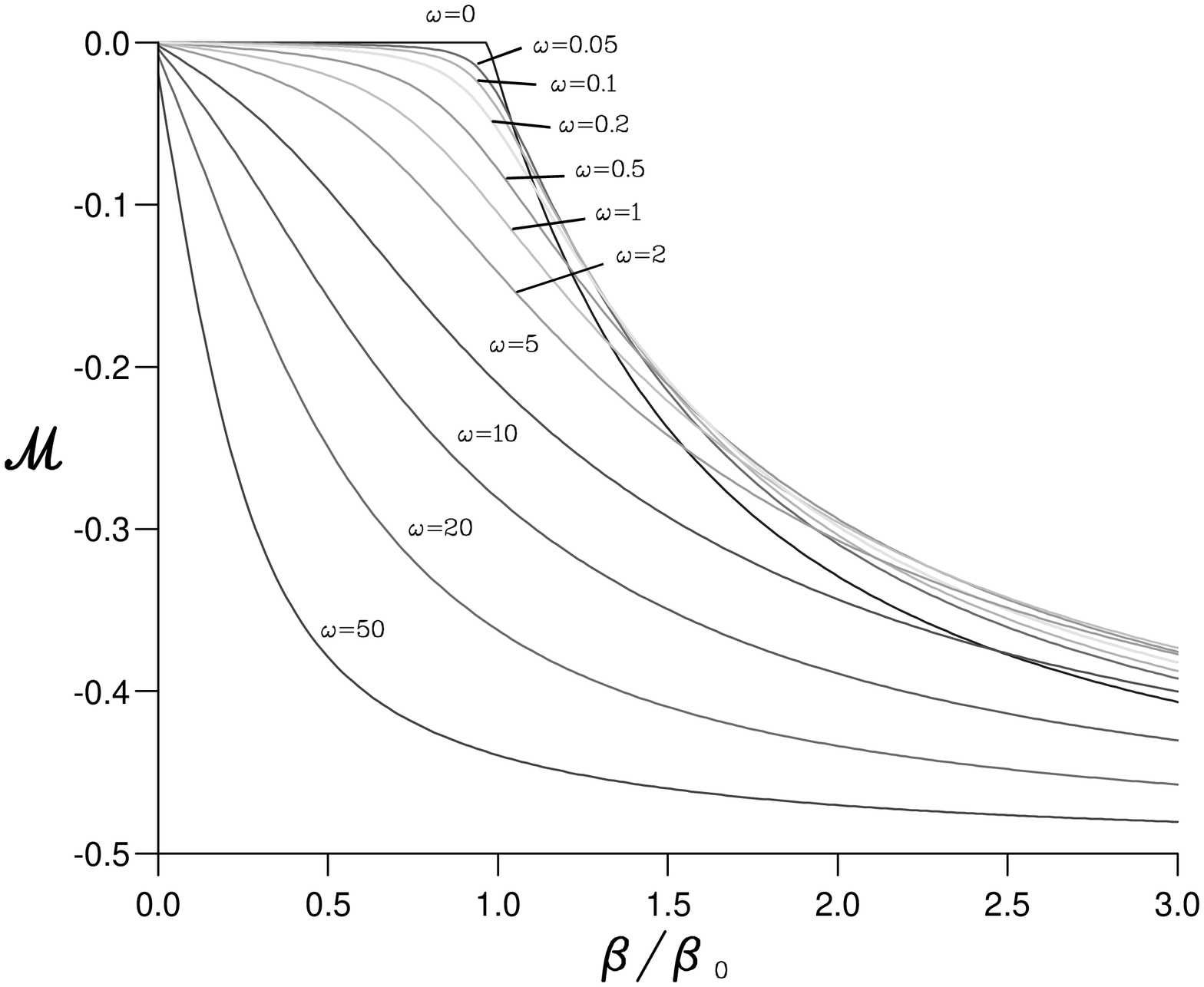}}
\setlength{\epsfxsize}{80mm}
\centerline{\epsfbox{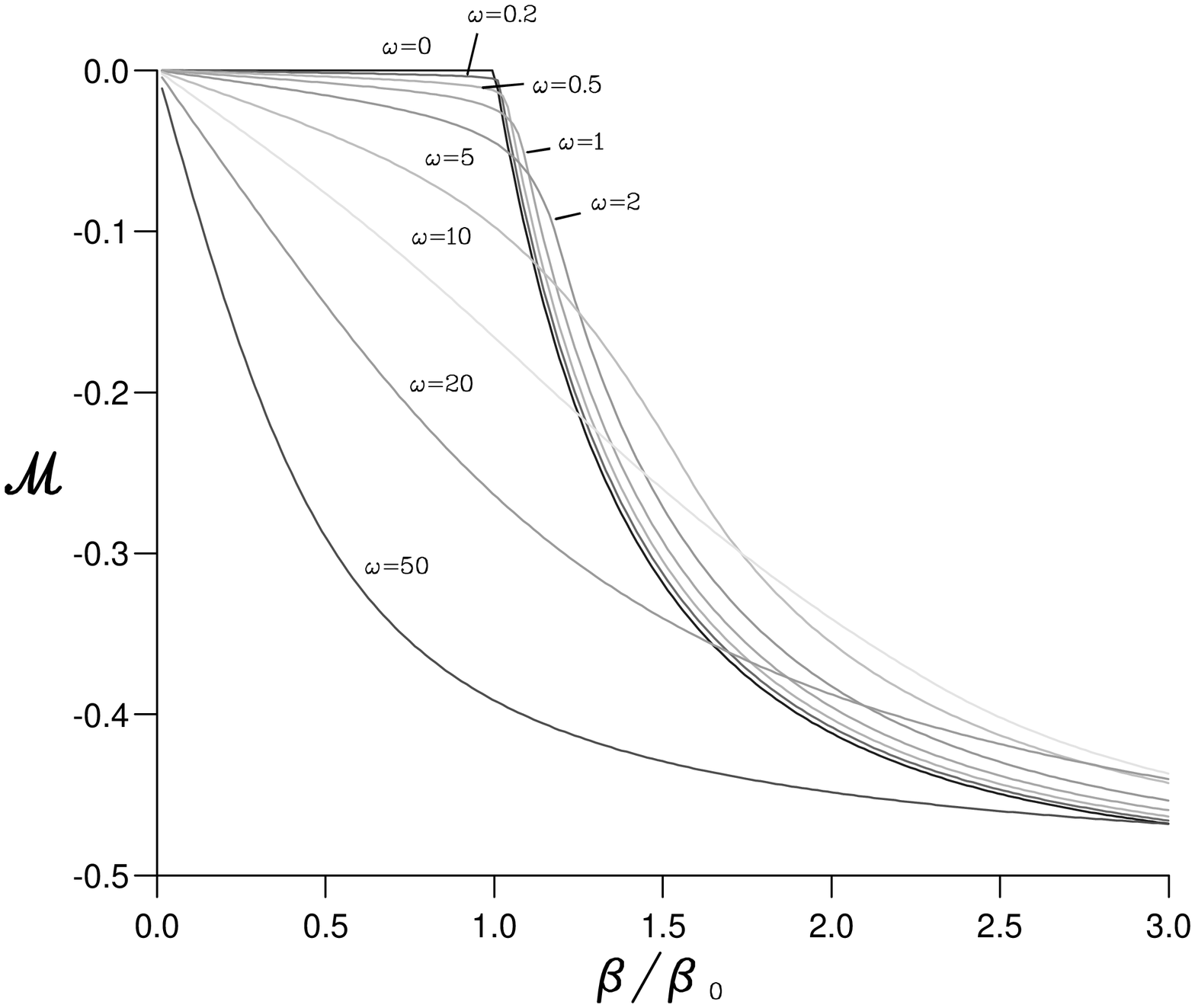}}
\setlength{\epsfxsize}{80mm}
\centerline{\epsfbox{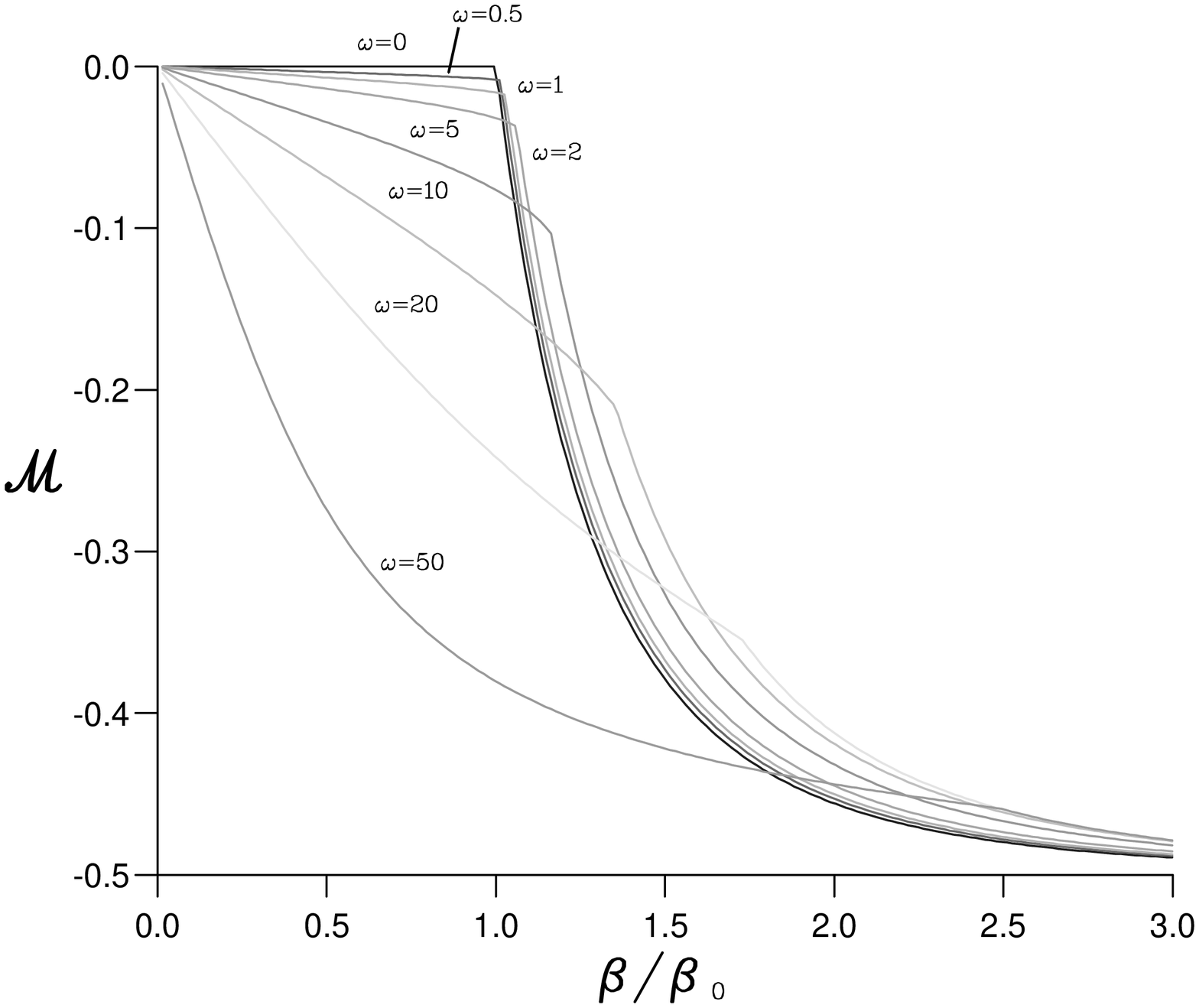}}
\caption{The dimensionless magnetization $\protect{{\cal M}=\frac{mVM}{Q}}$ for the charged bose gas in 3, 5, and 7 spatial 
dimensions. As the number of dimensions increases, the diamagnetism due to macroscopic occupation of the 
lowest Landau level in the low temperature (high $\frac{\beta}{\beta_{0}}$) region becomes more pronounced. 
As can be seen from the graphs, the general behavior of the magnetized gas is similar in cases when a phase 
transition is absent ($D=3$) or present ($D=5, 7$). The even dimensional cases ($D=4, 6$) have been omitted 
for brevity as they merely interpolate between the odd ones shown above.}
\label{fig2m}
\end{figure}

A selection of graphs for the dimensionless magnetization ${\cal M}=\frac{mVM}{Q}$ is shown in figure \ref{fig2m}. Although the 
expression for the magnetization will not be correct for $D\geq5$ if the condensate is neglected it can be seen 
that in the case $D=5$, the graph is smooth and continuous for all values of the field. For $D=7$ a kink is
apparent. This seeming incongruity can be explained by the fact that the magnetization is calculated from a 
first derivative of the thermodynamic potential whereas the heat capacity comes from a second derivative.
Hence the magnetization curve is smooth in those cases ($D=5,6$) where the heat capacity is continuous.

The zero--field spontaneous magnetization plotted on the graphs is of the Schafroth \cite{scha55a} form
\begin{equation}
\label{34150}
M=-\frac{Q}{2mV}\left[1-\left(\frac{T}{T_{0}}\right)^{\frac{D}{2}}\right]
\end{equation}
and it can be seen that as the field $B\rightarrow 0$, this limit is recovered.

\section{Analytical expansions}
\label{sec:meb}

So far we have presented mainly the results obtained from a numerical evaluations of the sums (\ref{32110}), 
since it is not possible to obtain exact results. However due to the form of the exponential involving $e^{-lx}$,
in order to obtain reliable numerical results directly from (\ref{32110}) it is necessary to include an 
increasing number of terms as $x$ decreases. In this section we will discuss a reliable method for obtaining 
approximate analytical expressions for various situations when $\varepsilon$ and $x$ are small by finding 
asymptotic expansions for $\bes{\kappa}{\alpha}{\delta}$.

The basic method we will use here involves the Mellin--Barnes contour integral representation for the 
exponential function:
\begin{equation}
\label{41010}
e^{-v} = \int_{c-i\infty}^{c+i\infty}\frac{d\theta}{2\pi i}\Gamma(\theta)v^{-\theta}.
\end{equation}
Here $c$ is a constant with $\Real c>0$ so that the contour lies to the right of the poles of the Gamma function.
This is basically equivalent to the method used by Robinson \cite{robi51a} to obtain asymptotic expansions
for the Bose--Einstein functions. It was used for the 3--dimensional magnetized gas by Daicic and Frankel 
\cite{daicfran96a}, and has been used to discuss Bose--Einstein condensation in a harmonic oscillator 
potential in reference \cite{kirstoms96c}. There are various ways in which (\ref{41010}) can be used to obtain 
expansions over a range of $x, \varepsilon$ as discussed in references \cite{kirstoms96c,toms97a}. We will 
content ourselves with the simplest presentation here.

The details of how (\ref{41010}) may be used for the class of sums (\ref{32110}) are given in Appendix 
\ref{app:meb} for the situations of interest to us. The basic method is to use (\ref{41010}) to convert 
(\ref{32110}) into a contour integral. The contour may be closed in the left hand side of the complex plane 
and the result evaluated by the residue theorem. The even and odd spatial dimensions differ somewhat
in the pole structure of the integrand. The net result is an asymptotic series for 
$\bes{\kappa}{\alpha}{\delta}$ which can be used to approximate the specific heat, magnetization and 
other thermodynamic quantities.

\subsection{Critical temperature}
\label{subsec:crittemp}

For $D\geq 5$ the magnetized Bose gas is characterized by a well defined critical temperature $T_{\sss C}$ 
which satisfies (\ref{33080}). It is not possible to evaluate $T_{\sss C}$ in closed form. However we can use 
our asymptotic expansion of $\bes{1}{2-D}{0}$ to obtain an approximate result for weak magnetic fields. We
have from Appendix \ref{app:meb}
\begin{equation}
\label{41020}
Q\simeq eV \left(\frac{m}{2\pi \beta_{\sss C}}\right)^{\frac{D}{2}}\left\{\zer{\frac{D}{2}}
+\frac{x_{\sss C}}{2} \zer{\frac{D-2}{2}}+\dots\right\}
\end{equation}
if only the two leading terms are included. This assumes $x_{\sss C}\ll 1$.
With the free Bose gas critical temperature defined by 
\begin{equation}
\label{41030}
Q= eV \left(\frac{m}{2\pi \beta_{0}}\right)^{\frac{D}{2}}\zer{\frac{D}{2}}
\end{equation}
we find
\begin{equation}
\label{41040}
0\simeq \left[1-\left(\frac{\beta_{\sss C}}{\beta_{0}}\right)^{\frac{D}{2}}\right]\zer{\frac{D}{2}}
+\frac{x_{\sss C}}{2} \zer{\frac{D-2}{2}}.
\end{equation}
Since $x_{\sss C}$ is assumed small we see that $\beta_{\sss C}\simeq \beta_{0}$. It is easy to show from
(\ref{41040}) that 
\begin{equation}
\label{41050}
T_{\sss C}\simeq T_{0} -\frac{1}{D}\frac{\zer{\frac{D-2}{2}}}{\zer{\frac{D}{2}}}\frac{eB}{m}
\end{equation}
to leading order in $\frac{eB}{m}$. This shows that $T_{\sss C} \rightarrow T_{0}$ as $B\rightarrow 0$. 
Furthermore, for a fixed charge density, the critical temperature is lower when a non zero magnetic field 
is present. This is consistent with our earlier numerical results.

A cruder estimate of $T_{\sss C}$ was given in reference \cite{toms95b} which had the same linear 
behavior as in (\ref{41050}) but with a different numerical factor in front of $B$. Our result (\ref{41050})
is a special case of the multi--component magnetic field presented in reference \cite{toms97a}.

It is possible to improve on the linear approximation of (\ref{41050}) by working consistently to higher order 
in the expansions. It is necessary to deal with $D=5, 6$ separately from $D>6$ because of the order of the 
terms retained. For $D=5$ we find
\begin{eqnarray}
\nn
\frac{T_{\sss C}}{T_{0}}&\simeq& 1- \frac{1}{5}\frac{\zer{\frac{3}{2}}}{\zer{\frac{5}{2}}} x_{0}
- \frac{1}{5\pi^{\frac{1}{2}}}\frac{\zer{\frac{3}{2}}}{\zer{\frac{5}{2}}} x_{0}^{\frac{3}{2}} \\
\label{41053}
&&+\left[\frac{\zertwo{\frac{3}{2}}}{100\zertwo{\frac{5}{2}}} 
- \frac{\zer{\frac{1}{2}}}{30\zer{\frac{5}{2}}} \right] x_{0}^{2} + {\cal O}\left(x_{0}^{\frac{5}{2}}\right).
\end{eqnarray}
For $D=6$ we find 
\begin{eqnarray}
\nn
\frac{T_{\sss C}}{T_{0}}&\simeq& 1- \frac{\pi^{2}}{36\zer{3}} x_{0} +\frac{x_{0}^{2}}{6\zer{3}} \\
\label{41056}
&&\times \left[\frac{\zeta'(2)}{\pi^{2}} + \frac{\pi^{4}}{216\zer{3}} 
+\frac{1}{6} \left(\ln{\frac{x_{0}}{2\pi}}-6 \right) \right]  + \dots.
\end{eqnarray}
For $D\geq 7$ we have
\begin{eqnarray}
\nn
\frac{T_{\sss C}}{T_{0}}&\simeq&1- \frac{1}{D}\frac{\zer{\frac{D-2}{2}}}{\zer{\frac{D}{2}}} x_{0} 
+\frac{1}{6D}\frac{\zer{\frac{D-2}{2}}}{\zer{\frac{D}{2}}}\\
\label{41059}
&&\times \left[\left(\frac{3(D-2)}{2D}\right)
\frac{\zer{\frac{D-2}{2}}}{\zer{\frac{D}{2}}}-1\right] x_{0}^{2} + \dots.
\end{eqnarray}
It is worth remarking that the result given by May \cite{may65a} is only correct if it is taken to 
linear order in $B$.

We can also obtain an approximate expression for the charge in the condensate when $T\leq T_{\sss C}$
for $D\geq 5$. When $T\leq T_{\sss C}$ we have $\varepsilon=0$, so (\ref{32130}) gives us
\begin{equation}
\label{41060}
Q_{1} = e V \left(\frac{m}{2 \pi \beta} \right)^{\frac{D}{2}} x \bes{1}{2-D}{0}|_{\varepsilon=0}.
\end{equation}
By using (\ref{33080}), which defines the critical temperature $T_{\sss C}$, we find 
(noting $x=\beta\omega$, $x_{\sss C}=\beta_{\sss C}\omega$)
\begin{equation}
\label{41070}
Q_{1} = Q \left(\frac{x_{\sss C}}{x}\right)^{\frac{D-2}{2}}
\frac{\bes{1}{2-D}{0}|_{\varepsilon=0}}{\bes{1}{2-D}{0}|_{\sstack{\varepsilon=0,}{x=x_{\sss C}}}}.
\end{equation}
From (\ref{31100}) we find that the charge in the condensate is 
\begin{equation}
\label{41080}
Q_{0} = Q \left\{1-\left(\frac{x_{\sss C}}{x}\right)^{\frac{D-2}{2}}
\frac{\bes{1}{2-D}{0}|_{\varepsilon=0}}{\bes{1}{2-D}{0}|_{\sstack{\varepsilon=0}{x=x_{\sss C}}}}\right\}.
\end{equation}
This result is exact. If we now use the approximate analytical expressions for $\Sigma_{1}$ we obtain
\begin{equation}
\label{41090}
Q_{0} = Q \left\{1-\left(\frac{x_{\sss C}}{x}\right)^{\frac{D}{2}}
-\frac{1}{2}\left(\frac{x_{\sss C}}{x}\right)^{\frac{D}{2}}
\frac{\zer{\frac{D-2}{2}}}{\zer{\frac{D}{2}}}(x-x_{\sss C})\right\}.
\end{equation}
It is worth remarking that the accuracy of any of our approximate expressions can be increased by simply 
including more terms.

From (\ref{41090}) we see that as $\omega\rightarrow 0$ because $\frac{x_{\sss C}}{x}=\frac{T}{T_{\sss C}}$,
and we know that $T_{\sss C}\rightarrow T_{0}$, we recover the free Bose gas result
\begin{equation}
\label{41100}
Q_{0}=Q\left[1-\left(\frac{T}{T_{0}}\right)^{\frac{D}{2}}\right]
\end{equation}
The term in (\ref{41090}) which involves $(x-x_{\sss C})$ represents the lowest order correction to the 
free field result in a weak magnetic field.

\begin{figure}
\setlength{\epsfxsize}{80mm}
\centerline{\epsfbox{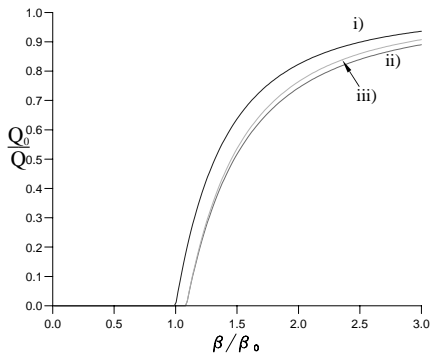}}
\caption{The ratio of the charge contained in the ground state over the total charge for i) the free gas 
(\ref{41100}), ii) the exact result (\ref{41080}), and iii) the approximation (\ref{41090}) (for $D=5$, $\omega=1$).}
\label{fig8}
\end{figure}

In figure \ref{fig8} we have plotted the free gas result (\ref{41100}), the exact numerical result, and our
approximation (\ref{41090}) for the case $D=5$. It can be seen that (\ref{41090}) is very close to the true result.

\subsection{Magnetization for $D\geq 5$}
\label{subsec:magdgeq5}

The magnetization may be evaluated using (\ref{34140}). It is necessary to distinguish the cases $D=3,4$ for
which no phase transition occurs from $D\geq 5$. We will deal with $D\geq 5$ in this section, leaving $D=3,4$ 
until later. With $D\geq 5$ we must deal with $T\geq T_{\sss C}$ and $T\leq T_{\sss C}$ since the chemical 
potential is different in the two cases. A further complication is that even and odd dimensions must be treated 
separately, as explained in Appendix \ref{app:meb}, and in addition the leading terms are different when 
$D=5,6$ than for $D>6$. For simplicity we will only give the expressions for $T\leq T_{\sss C}$. When 
$T\geq T_{\sss C}$ we have $\varepsilon\neq 0$ and the results are more complicated. In addition, the 
expansions break down once $\varepsilon$ grows too large.

For $T\leq T_{\sss C}$ we use (\ref{34140}) with $\varepsilon=0$: 
\begin{eqnarray}
\nn
M&=& -\frac{Q}{2mV}+\frac{e}{m} \left(\frac{m}{2 \pi \beta}\right)^{\frac{D}{2}} \left\{\bes{1}{-D}{0} 
\right. \\
\label{42010}
&& \left. - x \bes{2}{2-D}{1} \right\}|_{\varepsilon=0}.
\end{eqnarray}
This can be rewritten if we eliminate $\left(\frac{m}{2\pi}\right)^{\frac{D}{2}}$ using (\ref{33080}). We find
\begin{eqnarray}
\nn
M &=& -\frac{Q}{2mV}\left\{1 -2\left(\frac{T}{T_{\sss C}}\right)^{\frac{D}{2}}\right. \\
\label{42020}
&& \left. \times \frac{\left(\bes{1}{-D}{0} -x \bes{2}{2-D}{1}\right)|_{\varepsilon=0}}
{x_{\sss C} \bes{1}{2-D}{0}|_{\sstack{\varepsilon=0,}{x=x_{\sss C}}}}\right\}
\end{eqnarray}
It is now a straightforward matter to use the results of Appendix \ref{app:meb} to expand the sums.

\subsubsection{$D=5$}

From (\ref{A1110}) with $\alpha=-5$ and $\varepsilon=0$ we have
\begin{eqnarray}
\nn
\bes{1}{-5}{0} &\simeq& \frac{\zeta_{\sss R}\left(\frac{7}{2}\right)}{x} + \frac{\zer{\frac{5}{2}}}{2} 
+\frac{x \zer{\frac{3}{2}}}{12} \\
\label{42030}
&& +\frac{x^{\frac{3}{2}} \zer{\frac{5}{2}}}{4\pi^{\frac{3}{2}}} 
-\frac{x^{3}\zer{\frac{1}{2}}}{720} + {\cal O}\left(x^{5}\right)
\end{eqnarray}
From (\ref{A2070}) with $\alpha=-3$, $\varepsilon=0$ and $\delta=1$ we have
\begin{eqnarray}
\nn
\bes{2}{-3}{1}&\simeq& \frac{\zeta_{\sss R}\left(\frac{7}{2}\right)}{x^{2}} 
-\frac{\zer{\frac{3}{2}}}{12} -\frac{3 x^{\frac{1}{2}} \zer{\frac{5}{2}}}{8\pi^{\frac{3}{2}}} \\
\label{42040}
&& -\frac{x^{2}\zer{\frac{3}{2}}}{960\pi} +{\cal O}\left(x^{4}\right)
\end{eqnarray}
From (\ref{A1110}) with $\alpha=-3$ and $\varepsilon=0$ we have
\begin{eqnarray}
\nn
\bes{1}{-3}{0}|_{x=x_{\sss C}} &\simeq& \frac{\zeta_{\sss R}\left(\frac{5}{2}\right)}{x_{\sss C}} 
+ \frac{\zer{\frac{3}{2}}}{2} 
+\frac{x_{\sss C}^{\frac{1}{2}} \zeta_{\sss R}\left(\frac{3}{2}\right)}{2\pi^{\frac{1}{2}}} \\
\label{42050}
&&+ \frac{\zer{\frac{1}{2}}}{12} x_{\sss C} + \frac{x_{\sss C}^{3} \zer{\frac{3}{2}}}{2880 \pi} 
+ {\cal O}\left(x_{\sss C}^{5}\right)
\end{eqnarray}
It is now easy to show that the asymptotic expansion of the magnetization is given by
\begin{equation}
\label{42060}
M=-\frac{Q}{2m V}\left\{1-\left(\frac{T}{T_{\sss C}}\right)^{\frac{5}{2}}
-\left(\frac{T}{T_{\sss C}}\right)^{\frac{5}{2}}f_{5}\left(x,x_{\sss C}\right)\right\}
\end{equation}
where
\begin{eqnarray}
\nn
f_{5}\left(x,x_{\sss C}\right)&\simeq&
\left(\frac{x}{3}-\frac{x_{\sss C}}{2}\right)\frac{\zer{\frac{3}{2}}}{\zer{\frac{5}{2}}}
+\frac{5 x^{\frac{3}{2}}}{4\pi^{\frac{3}{2}}}-\frac{x_{\sss C}^{\frac{3}{2}}}{2\pi^{\frac{1}{2}}}
\frac{\zer{\frac{3}{2}}}{\zer{\frac{5}{2}}} \\
\nn
&& +\left(\frac{x_{\sss C}^{2}}{4}-\frac{x x_{\sss C}}{6}\right)
\frac{\zertwo{\frac{3}{2}}}{\zertwo{\frac{5}{2}}}
-\frac{x_{\sss C}^{2} \zer{\frac{1}{2}}}{12 \zer{\frac{5}{2}}} \\
\nn
&&+\left(x_{\sss C}-\frac{x}{3}\right)\frac{x_{\sss C}^{\frac{3}{2}}}{2\pi^{\frac{1}{2}}}
\frac{\zertwo{\frac{3}{2}}}{\zertwo{\frac{5}{2}}} \\
\label{42065}
&&-\frac{5x_{\sss C} x^{\frac{3}{2}}}{8\pi^{\frac{3}{2}}}\frac{\zer{\frac{3}{2}}}{\zer{\frac{5}{2}}}
+\dots
\end{eqnarray}

By taking $B\rightarrow 0$ we are left with the spontaneous magnetization
\begin{equation}
\label{42070}
M(B\rightarrow 0)\simeq -\frac{Q}{2m V} \left[1-\left(\frac{T}{T_{0}}\right)^{\frac{5}{2}}\right]
\end{equation}
since $T_{\sss C}\rightarrow T_{0}$ and $f_{5}\left(x,x_{\sss C}\right)\rightarrow 0$ in this limit. 
This is the 5--dimensional version of Schafroth's result, and agrees with May \cite{may65a}. 
When $B\neq 0$ there are corrections to the Schafroth form as shown by the terms in 
$f_{5}\left(x,x_{\sss C}\right)$. There are two sources for these corrections. The first is that for $B\neq 0$, 
$T_{\sss C}\neq T_{0}$. The second is that the asymptotic form of $M$ has higher order terms present.
The fact that $M$ does not vanish as $B\rightarrow 0$ shows that the Meissner--Ochsenfeld effect exists.

We can also compare our result with that given by May \cite{may65a}. To do this we must replace 
$T_{\sss C}$ in (\ref{42060}) with $T_{0}$. We will only work to 
first order in the magnetic field and use (\ref{42060}). It is easily shown that
\begin{eqnarray}
\nn
M&\simeq &-\frac{Q}{2m V} \left\{1-\left(\frac{T}{T_{0}}\right)^{\frac{5}{2}}\right. \\
\label{42080}
&& \left. -\frac{1}{3}\left(\frac{T}{T_{0}}\right)^{\frac{5}{2}}\frac{\zer{\frac{3}{2}}}{\zer{\frac{5}{2}}}x
+\dots\right\}
\end{eqnarray}
This agrees with May's result where a different method was used. There are two comments to make here.
The first is that had we taken the expansion beyond linear order in $x$ the results obtained would not
agree with that of May's method for reasons already mentioned. Secondly, care must be exercised 
in using (\ref{42080}) because we have assumed $T\leq T_{\sss C}$ in its derivation. This means that 
(\ref{42080}) does not hold at $T=T_{0}$, as can be seen from \ref{41050}. If we set $T=T_{0}$ 
in (\ref{42080}) we would conclude that $M$ was positive, leading to paramagnetic rather than diamagnetic 
behavior. This is clearly wrong. We will return to results for $T=T_{0}$ later. There is of course nothing 
wrong with taking $T=T_{\sss C}$ in (\ref{42060}).

\subsubsection{$D=6$}

From (\ref{A1080}), (\ref{A1090}) with $\varepsilon=0$, $k=2$ we have
\begin{eqnarray}
\nn
\bes{1}{-4}{0}|_{\sstack{\varepsilon=0,}{x=x_{\sss C}}} &\simeq& 
\frac{\zer{3}}{x_{\sss C}} +\frac{\zer{2}}{2} \\
\nn
&&-x_{\sss C} \left[\frac{{\zeta}_{\sss R}'(2)}{2\pi^{2}}
+\frac{(\ln{\frac{x_{\sss C}}{2\pi}}-\gamma)}{12}\right] \\
\label{42090}
&&+ \frac{x_{\sss C}^{3}}{8640}+{\cal O}\left(x_{\sss C}^{5}\right)
\end{eqnarray}
From (\ref{A1080}) with $\varepsilon=0$, $k=3$ we find
\begin{eqnarray}
\nn
\bes{1}{-6}{0} &\simeq& \frac{\zer{4}}{x} +\frac{\zer{3}}{2} +\frac{x \zer{2}}{12} \\
\label{42100}
&&-\frac{x^{2}\zer{3}}{8\pi^{2}} + \frac{x^{3}}{1440}.
\end{eqnarray}
In this case all higher order terms in the asymptotic expansion vanish. Finally from (\ref{A2060}) with 
 $\varepsilon=0$, $\delta=1$ and $k=2$ we find
\begin{eqnarray}
\label{42110}
\bes{2}{-4}{1}&\simeq& \frac{\zeta_{\sss R}\left(4\right)}{x^{2}} - \frac{\zer{2}}{12} 
+ \frac{x\zer{3}}{4\pi^{2}}-\frac{x^{2}}{480}. 
\end{eqnarray}
Again higher order terms in the asymptotic expansion vanish. In obtaining (\ref{42090}---\ref{42110}) 
various relations of the Riemann $\zeta$--function were used. The expression for the
magnetization becomes
\begin{equation}
\label{42120}
M=-\frac{Q}{2m V}\left\{1-\left(\frac{T}{T_{\sss C}}\right)^{3}
-\left(\frac{T}{T_{\sss C}}\right)^{3}f_{6}\left(x,x_{\sss C}\right)\right\}
\end{equation}
where
\begin{eqnarray}
\nn
f_{6}\left(x,x_{\sss C}\right)&\simeq&
-\left(\frac{x}{3}-\frac{x_{\sss C}}{2}\right)\frac{\pi^{2}}{6\zer{3}} \\
\nn
&&+\frac{3 x^{2}}{4\pi^{2}}+\frac{\pi^{4} x x_{\sss C}}{216\zertwo{3}}
-\frac{\pi^{4} x^{2}_{\sss C}}{144\zertwo{3}} \\
\nn
&& -\frac{x_{\sss C}^{2}}{\zer{3}}\left[\frac{\zeta'_{\sss R}(2)}{2\pi^{2}}
+\frac{\ln{\frac{x_{\sss C}}{2\pi}}}{12}-\frac{\gamma}{12}\right] \\
\nn
&& +x_{\sss C}^{3}\left[\frac{\pi^{6}}{1728\zeta^{3}_{\sss R}(3)} \right. \\
\nn
&&\left.+\frac{\pi^{2}}{6\zeta^{2}_{\sss R}(3)}
\left(\frac{\zeta'_{\sss R}(2)}{2\pi^{2}}+\frac{\ln{\frac{x_{\sss C}}{2\pi}}}{12}
-\frac{\gamma}{12}\right) \right] \\
\nn
&& -x x_{\sss C}^{2}\left[\frac{\pi^{6}}{2592\zeta^{3}_{\sss R}(3)}\right. \\
\nn
&&\left. +\frac{\pi^{2}}{18\zeta^{2}_{\sss R}(3)}
\left(\frac{\zeta'_{\sss R}(2)}{2\pi^{2}}+\frac{\ln{\frac{x_{\sss C}}{2\pi}}}{12}
-\frac{\gamma}{12}\right) \right] \\
\label{42130}
&&-\frac{x^{2}}{4\zer{3}}\left(\frac{x}{45}+\frac{x_{\sss C}}{4}\right)+\dots.
\end{eqnarray}
Again we see the Meissner--Ochsenfeld effect as $B\rightarrow 0$, with the magnetization of the 
Schafroth form.

\subsubsection{$D\geq 7$}

Using the results given in Appendix \ref{app:meb}, and working only to quadratic order in $x$, $x_{\sss C}$ 
(to avoid treating even and odd dimensions separately) we find
\begin{equation}
\label{42140}
M=-\frac{Q}{2mV}\left\{1-\left(\frac{T}{T_{\sss C}}\right)^{\frac{D}{2}}
-\left(\frac{T}{T_{\sss C}}\right)^{\frac{D}{2}} f_{\sss D}\left(x,x_{\sss C}\right)\right\}
\end{equation}
where
\begin{eqnarray}
\nn
f_{\sss D}\left(x,x_{\sss C}\right)&\simeq&
\left(\frac{x}{3}-\frac{x_{\sss C}}{2}\right)\frac{\zer{\frac{D-2}{2}}}{\zer{\frac{D}{2}}} \\
\nn
&&+\frac{x_{\sss C}^{2}}{4}\left[\frac{\zeta_{\sss R}^{2}\left(\frac{D-2}{2}\right)}
{\zeta_{\sss R}^{2}\left(\frac{D}{2}\right)}-\frac{\zer{\frac{D-2}{2}}}{3\zer{\frac{D}{2}}}\right]  \\
\label{42150}
&&- x x_{\sss C} \frac{\zeta_{\sss R}^{2}\left(\frac{D-2}{2}\right)}
{6\zeta_{\sss R}^{2}\left(\frac{D}{2}\right)}+\dots.
\end{eqnarray}
Since $f_{\sss D}\left(x,x_{\sss C}\right)\rightarrow 0$ as $B\rightarrow 0$ we recover the Schafroth 
magnetization.

\subsection{Specific Heat}
\label{subsec:cvexp}

In a similar manner to the previous section analytical expansions may be obtained for the specific heat 
capacities in the dimensions of interest to us in this paper. It is more convenient to express these results in
the form they were given previously, {\it i.e.} as expansions of the ratio $\frac{C_{v}}{N}$.
These expansions are constructed from the expression for the heat capacity (either equation (\ref{33110}) 
if above the transition temperature, or (\ref{33120}) if below it)  as well as that for the number density 
(obtained from (\ref{32130}) as $N=\frac{Q}{e}$) . The various sums present in these equations are 
treated as before by substituting the appropriate analytic approximations given in appendix \ref{app:meb}. 
Brief comments on the form of the expansions are made in each case.

\subsubsection{$D=3$}

From figure \ref{fig2} it can be seen that the charged Bose gas in three spatial 
dimensions does not exhibit a phase transition with finite field, but approaches the zero field result in the 
$B\rightarrow 0$ limit. The maximum of the heat capacity is always lower than the zero field limit, a fact 
which can clearly be seen from the analytic expansion
\begin{eqnarray}
\nn
\frac{C_{v}}{N}&\simeq&\frac{15\zeta_{\sss R}\left(\frac{5}{2}\right)}{4\zer{\frac{3}{2}}}
-\frac{3x^{\frac{1}{2}}}{4\sqrt{\pi} \zeta_{\sss R}^{2}\left(\frac{3}{2}\right) \zeh{\frac{3}{2}}{\varepsilon}}
\left[6 \zerthree{\frac{3}{2}} \right. \\
\nn
&&\left. +5\pi \zer{\frac{5}{2}}\zeh{\frac{1}{2}}{\varepsilon} \zeh{\frac{3}{2}}{\varepsilon}\right] \\
\nn
&& +\frac{3x}{8\pi\zeta_{\sss H}^{2}\left(\frac{3}{2},\varepsilon\right)
\zeta_{\sss R}^{2}\left(\frac{3}{2}\right)} 
\left[(1-2\varepsilon)\pi\zehtwo{\frac{3}{2}}{\varepsilon}
\zertwo{\frac{3}{2}} \right. \\
\nn
&& +5\pi(2\varepsilon-1)\zehtwo{\frac{3}{2}}{\varepsilon}
\zer{\frac{5}{2}}\zer{\frac{1}{2}}-24\zerthree{\frac{3}{2}}\zer{\frac{1}{2}} \\
\label{43010}
&&\left. -4\pi\zertwo{\frac{3}{2}}\zeh{\frac{1}{2}}{\varepsilon}
\zeh{\frac{3}{2}}{\varepsilon}\right] +{\cal O}(x^{\frac{3}{2}}).
\end{eqnarray}

\subsubsection{$D=4$}

In even dimensions, as noted in the appendix, the analysis is complicated by the presence of double poles.
These give rise to terms in $\ln{x}$ which dominate over the higher power terms which are the only ones 
present in the odd dimensional cases. In addition, the situation is complicated by terms involving 
$\ln{\varepsilon}$ and $\psi(\varepsilon)$. The expansion is given by
\begin{eqnarray}
\nn
\frac{C_{v}}{N}&\simeq&\frac{36\zeta_{\sss R}\left(3\right)}{\pi^{2}}
+\frac{2\pi^{2}\left(\ln{x}-\psi(\varepsilon)\right)}{3(\ln{x})^{2}}  \\
\nn
&& +\left[1-2\varepsilon+\frac{216\zer{3}}{\pi^{4}}
\ln{\varepsilon}+\zeta'_{\sss H}\left(0,1+\varepsilon\right) \right]x \\
\label{43020}
&&+{\cal O}\left(\frac{x}{\ln{x}}\right).
\end{eqnarray}

\subsubsection{$D=5$}

The five dimensional gas is the first to show the existence of a phase transition and as has previously 
been discussed in subsection \ref{subsec:cv}, it is necessary to use different expressions for the heat 
capacity above and below the critical temperature. For $T>T_{\sss C}$, this quantity is given by 
\begin{eqnarray}
\nn
\frac{C_{v}}{N}&\simeq&\frac{35\zeta_{\sss R}\left(\frac{7}{2}\right)\zer{\frac{3}{2}}
-25\zeta_{\sss R}^{2}\left(\frac{5}{2}\right)}{4\zer{\frac{3}{2}}\zer{\frac{5}{2}}} \\
\nn
&&+\frac{25 \pi^{\frac{1}{2}} x^{\frac{1}{2}}\zeh{\frac{1}{2}}{\varepsilon}
\zeta_{\sss R}\left(\frac{5}{2}\right)}{4\zeta_{\sss R}^{2}\left(\frac{3}{2}\right)} \\
\nn
&& +\frac{5x(2\varepsilon-1)}{8\zeta_{\sss R}^{2}\left(\frac{3}{2}\right)
\zeta_{\sss R}^{2}\left(\frac{5}{2}\right)} \left[7\zerthree{\frac{3}{2}} \zer{\frac{7}{2}} 
-5\zer{\frac{1}{2}}\zerthree{\frac{5}{2}} \right. \\
\label{43030}
&& \left.  -2\zertwo{\frac{3}{2}}\zertwo{\frac{5}{2}}\right] +{\cal O}(x^{\frac{3}{2}}),
\end{eqnarray}
while below $T_{\sss C}$ equation (\ref{33120}) must be used, and hence
\begin{eqnarray}
\nn
\frac{C_{v}}{N}&\simeq&\frac{35\zeta_{\sss R}\left(\frac{7}{2}\right)}{4\zer{\frac{5}{2}}} 
+\frac{x\left(15\zeta_{\sss R}^{2}\left(\frac{5}{2}\right)
-35\zer{\frac{3}{2}}\zer{\frac{7}{2}}\right)}
{8\zeta_{\sss R}^{2}\left(\frac{5}{2}\right)} \\
\nn
&& +\frac{35\pi^{\frac{1}{2}}x^{\frac{3}{2}}\zer{\frac{7}{2}}\zer{-\frac{1}{2}}}
{2\zeta_{\sss R}^{2}\left(\frac{5}{2}\right)} \\
\nn
&&-\frac{7x^{2}\left(6\zer{\frac{3}{2}}\zer{\frac{5}{2}}+5\zer{\frac{1}{2}}\zer{\frac{7}{2}}\right)}
{48\zeta_{\sss R}^{2}\left(\frac{5}{2}\right)}\\
\label{43040}
&&+{\cal O}(x^{3}).
\end{eqnarray}

\subsubsection{$D=6$}

The $D=6$ case is similar to the $D=4$ case except that here too a phase transition occurs below a 
critical temperature. Again the presence of logarithmic terms complicates the expression. For $T>T_{\sss C}$
\begin{eqnarray}
\nn
\frac{C_{v}}{N}&\simeq&\frac{2\left(\pi^{6}-405\zeta_{\sss R}^{3}\left(3\right)\right)}{15\pi^{2}\zer{3}} 
+ \frac{x}{90\pi^{4}\zeta_{\sss R}^{2}(3)} \\
\nn
&&\times\left[\pi^{4}(2\varepsilon-1)\left(\pi^{6}-135\zeta_{\sss R}^{2}(3)\right) \right. \\
\nn
&& \left. +29160\zeta_{\sss R}^{3}(3)\left(\zeta'_{\sss H}(0,1+\varepsilon)-\ln{\varepsilon}\right)\right] \\
\nn
&&+\frac{162(2\varepsilon-1)x\ln{x}}{\pi^{4}} \\
\label{43050}
&&+{\cal O}\left(x^{2}\right),
\end{eqnarray}
whilst for $T\leq T_{\sss C}$
\begin{eqnarray}
\nn
\frac{C_{v}}{N}&\simeq&\frac{2\pi^{4}}{15\zer{3}}
+\frac{x\left(270\zeta_{\sss R}^{2}(3)-\pi^{6}\right)}{90 \zeta_{\sss R}^{2}(3)} 
+ \frac{\pi^{2}x^{2}}{180 \zeta_{\sss R}^{2}(3)} \\
\nn
&& \times \left[24\pi^{2}\zeta'_{\sss H}\left(-1,1\right) -2\pi^{2}-125 \zer{3}\right] \\
\label{43060}
&&+{\cal O}\left(x^{2}\ln{x}\right).
\end{eqnarray}

\subsubsection{$D=7$}

$D=7$ is the first case in which the heat capacity is wholly discontinuous. Above the critical temperature
the expression
\begin{eqnarray}
\nn
\frac{C_{v}}{N}&\simeq&\frac{7\left(9\zeta_{\sss R}\left(\frac{5}{2}\right)\zer{\frac{9}{2}}
-7\zeta_{\sss R}^{2}\left(\frac{7}{2}\right)\right)}{4\zer{\frac{5}{2}}\zer{\frac{7}{2}}} \\
\nn
&& +\frac{x(14\varepsilon-7)}{8\zer{\frac{5}{2}}\zeta_{\sss R}^{2}\left(\frac{7}{2}\right)}
\left[5\zertwo{\frac{7}{2}} \zer{\frac{5}{2}} \right. \\
\nn
&&\left. +9\zer{\frac{3}{2}}\zer{\frac{5}{2}}\zer{\frac{9}{2}} 
-7\zertwo{\frac{7}{2}}\left\{\zer{\frac{3}{2}}+\zer{\frac{7}{2}}\right\}\right] \\
\label{43070}
&&-\frac{49 \pi^{\frac{1}{2}} x^{\frac{3}{2}}\zeh{-\frac{1}{2}}{\varepsilon}
\zeta_{\sss R}\left(\frac{7}{2}\right)}{2\zeta_{\sss R}^{2}\left(\frac{5}{2}\right)}
+{\cal O}(x^{2}),
\end{eqnarray}
holds whereas below it
\begin{eqnarray}
\nn
\frac{C_{v}}{N}&\simeq&\frac{63\zeta_{\sss R}\left(\frac{9}{2}\right)}{4\zer{\frac{7}{2}}} 
+\frac{7x\left(5\zeta_{\sss R}^{2}\left(\frac{7}{2}\right) -9\zer{\frac{3}{2}}\zer{\frac{9}{2}}\right)}
{8\zeta_{\sss R}^{2}\left(\frac{7}{2}\right)} \\
\nn
&& +\frac{x^{2}}{16\zeta_{\sss R}^{2}\left(\frac{7}{2}\right)}
\left[5\zer{\frac{7}{2}}\zer{\frac{5}{2}}-35\zer{\frac{3}{2}}\zer{\frac{7}{2}} \right. \\
\label{43080}
&& \left. -21\zer{\frac{1}{2}}\zer{\frac{9}{2}}\right]+{\cal O}(x^{3}).
\end{eqnarray}
must be used.

\subsection{Expansions for $T$ near $T_{0}$}
\label{subsec:expan}

As noted in Schafroth's original paper \cite{scha55a} the approximation (for the 3-dimensional gas)
\begin{equation}
\label{44010}
M\simeq -\frac{Q}{2mV}\left[1-\left(\frac{T}{T_{0}}\right)^{\frac{3}{2}}\right]
\end{equation}
will break down when $T$ becomes too close to $T_{0}$. This can be substantiated by a direct numerical 
evaluation of $M$ and comparison with (\ref{44010}) as we showed in \cite{stantoms97a}. The Schafroth 
criterion for validity of (\ref{44010}) can be derived in a simple manner as described in \cite{daicfran96a}. 
An obvious question to ask is that if (\ref{44010}) does not hold as $T$ becomes close to $T_{0}$, is there 
another simple approximation which can be used? This was first studied by Daicic and Frankel 
\cite{daicfran96a}, who showed by expanding about $\mu=0$ that the magnetization 
$M\simeq - c B^{\frac{1}{2}}$ with $c>0$ a constant. This approximation was valid for values of $T$ closer to 
$T_{0}$ than Schafroth's approximation, but still broke down as $T\rightarrow T_{0}$. In reference 
\cite{stantoms97a} we showed how it was possible to evaluate the magnetization in a temperature range 
which included $T=T_{0}$. In this section we will present the details of this result and generalize the 
analysis to spatial dimensions $D>3$.

For $D=3$ the critical temperature $T_{0}$ for the free Bose gas is defined by
\begin{equation}
\label{44020}
Q=eV\left(\frac{m}{2\pi\beta_{0}}\right)^{\frac{3}{2}}\zer{\frac{3}{2}}
\end{equation}
where $\beta_{0}=T_{0}^{-1}$. For $B\neq 0$ we have 
\begin{equation}
\label{44030}
Q=eV\left(\frac{m}{2\pi\beta}\right)^{\frac{3}{2}}x\bes{1}{-1}{0}.
\end{equation}
Equating these two expressions gives
\begin{equation}
\label{44040}
x\bes{1}{-1}{0}=\left(\frac{x}{x_{0}}\right)^{\frac{3}{2}}\zer{\frac{3}{2}}
\end{equation}
where $x_{0}=\beta_{0} \omega$. The aim now is to solve this for $\varepsilon$ when the magnetic field is 
weak and $T$ is close to $T_{0}$ (meaning $x$ is close to $x_{0}$). Because we assume a weak magnetic field, 
we may use the first few terms in the asymptotic expansion of $\bes{1}{-1}{0}$ obtained from Appendix 
\ref{app:meb}. We will approximate (\ref{44040}) as
\begin{eqnarray}
\nn
\left(\frac{x}{x_{0}}\right)^{\frac{3}{2}}\zer{\frac{3}{2}}&\simeq& \zer{\frac{3}{2}}
+ (\pi x)^{\frac{1}{2}} \zeh{\frac{1}{2}}{\varepsilon} \\
\label{44050}
&&+\zer{\frac{1}{2}}\left(\frac{1}{2}-\varepsilon\right)x + \dots.
\end{eqnarray}
Higher order terms could easily be included to improve the accuracy of the result.

Suppose that we concentrate first on $T=T_{0}$. Call the value of $\varepsilon$ at $T=T_{0}$, 
$\varepsilon_{0}$. Then (\ref{44050}) gives us 
\begin{equation}
\label{44060}
0=\zeh{\frac{1}{2}}{\varepsilon_{0}}+\pi^{-\frac{1}{2}}\zer{\frac{1}{2}}
\left(\frac{1}{2}-\varepsilon_{0}\right)x_{0}^{\frac{1}{2}} + \dots
\end{equation}
(The next term is of order $x^{\frac{3}{2}}$). If we let $B\rightarrow 0$ then 
$x_{0}=\beta_{0}\omega\rightarrow 0$. Thus as $B\rightarrow 0$ we must have 
$\varepsilon_{0}\rightarrow a$ where $a$ is defined by
\begin{equation}
\label{44070}
0=\zeh{\frac{1}{2}}{a}.
\end{equation}
The value of $a$ can be found numerically with the result $a=0.302721829\dots$. We have verified this by solving
(\ref{44040}) numerically for decreasing values of $B$ and found that $\varepsilon\rightarrow a$ as $B$ is 
reduced. Because $\mu=\omega\left(\frac{1}{2}-\varepsilon\right)$ this result is still consistent with the 
expectation that $\mu\rightarrow 0$ as $B\rightarrow 0$. We have essentially determined how fast $\mu$ 
vanishes as $B\rightarrow 0$.

For small, but non--zero, values of $B$ we can try to solve (\ref{44060}). In order to obtain a consistent 
expansion from (\ref{44060}) it is fairly clear that we must have
\begin{equation}
\label{44080}
\varepsilon_{0}\simeq a +a_{1} x_{0}^{\frac{1}{2}} + a_{2} x_{0} + \dots
\end{equation}
for some coefficients $a_{1},a_{2},\dots$ which can be found by substituting (\ref{44080}) into (\ref{44060})
and working to a consistent order in $x_{0}$. It is easily shown that
\begin{equation}
\label{44090}
\zeh{\frac{1}{2}}{\varepsilon_{0}}\simeq -\frac{1}{2}a_{1}x_{0}^{\frac{1}{2}}\zeh{\frac{3}{2}}{a}
+{\cal O}\left(x_{0}\right).
\end{equation}
Use of (\ref{44080}) and (\ref{44090}) in (\ref{44060}) fixes
\begin{equation}
\label{440100}
a_{1}=\frac{2\zer{\frac{1}{2}}}{\pi^{\frac{1}{2}}\zeh{\frac{3}{2}}{a}}\left(\frac{1}{2}-a\right).
\end{equation}
It should be clear how we can obtain an approximation for $\varepsilon_{0}$ to any order in $x_{0}$ by 
extending the procedure we have just described to higher order.

So far we have just concentrated on the evaluation of $\varepsilon$ at the single temperature $T_{0}$.
Suppose that we now extend this to temperatures which are close to $T_{0}$. By a simple extension of 
the analysis just presented it is possible to show
\begin{equation}
\label{440110}
\varepsilon\simeq a+ a_{1} x_{0}^{\frac{1}{2}}+
\frac{6\zer{\frac{3}{2}}}{\pi^{\frac{1}{2}}\zeh{\frac{3}{2}}{a}}x_{0}^{-\frac{1}{2}}
\left[1-\left(\frac{x}{x_{0}}\right)^{\frac{1}{2}}\right]+\dots.
\end{equation}
Consistency of this expansion requires
\begin{equation}
\label{440120}
1-\left(\frac{x}{x_{0}}\right)^{\frac{1}{2}}\ll x_{0}^{\frac{1}{2}}.
\end{equation}
In particular, the approximation is good at $x=x_{0}$.

The approximation we have found for $\varepsilon$ may now be used in the expressions for the specific 
heat and the magnetization. For the specific heat at $T=T_{0}$ we find
\begin{eqnarray}
\nn
\frac{C_{v}}{N}&\simeq&\frac{15\zer{\frac{5}{2}}}{4\zer{\frac{3}{2}}} 
- x_{0}^{\frac{1}{2}} \frac{9\zer{\frac{3}{2}}}{2\pi^{\frac{1}{2}}\zeh{\frac{3}{2}}{a}} \\
\nn
&&+x_{0}\left\{\frac{3}{4}\left(\frac{1}{2}-a\right)
+\frac{9\zer{\frac{1}{2}}\zer{\frac{3}{2}}}{\pi \zeta_{\sss H}^{2}\left(\frac{3}{2},a\right)} \right. \\
\nn
&&\left.-\frac{27 \zer{\frac{1}{2}}\zer{\frac{3}{2}}\zeh{\frac{5}{2}}{a}}
{2\pi \zeta_{\sss H}^{3}\left(\frac{3}{2},a\right)}\left(\frac{1}{2}-a\right)\right\} \\
\label{44130}
&&+{\cal O}\left(x_{0}^{\frac{3}{2}}\right)
\end{eqnarray}
This shows that as $B\rightarrow 0$ the specific heat approaches the free field result at $T=T_{0}$, 
confirming analytically the trend we found numerically. It also provides an analytic proof that for small 
$x_{0}$, the presence of the magnetic field lowers the value of the specific heat from the free field value.

For the magnetization we find (at $T=T_{0}$)
\begin{eqnarray}
\nn
M&\simeq&-\frac{Q}{2mV}\left\{\frac{6\pi^{\frac{1}{2}}\zeh{-\frac{1}{2}}{a}}{\zer{\frac{3}{2}}}
x_{0}^{\frac{1}{2}} \right. \\
\label{440140}
&&\left.-2\frac{\zer{\frac{1}{2}}}{\zer{\frac{3}{2}}}\left(\frac{1}{6}-a+a^{2}\right)x_{0}+\dots\right\}
\end{eqnarray}
This provides confirmation of the $M\simeq -c B^{\frac{1}{2}}$ magnetization law of Daicic and Frankel 
\cite{daicfran96a}, but at a lower temperature (and therefore the coefficient of $B^{\frac{1}{2}}$ differs 
from that of \cite{daicfran96a}). In addition, we have computed the next order correction to the leading 
$B^{\frac{1}{2}}$ behavior.

Having presented the case $D=3$ in some detail, we will now examine other dimensions in less detail. For 
$D=4$, in place of (\ref{44040}) we have
\begin{equation}
\label{44150}
x\bes{1}{-2}{0}= \left(\frac{x}{x_{0}}\right)^{2}\zer{2},
\end{equation}
with $T_{0}$ now defined by 
\begin{equation}
\label{44160}
Q=eV \left(\frac{m}{2\pi \beta_{0}}\right)^{2}\zer{2}.
\end{equation}
(Here $x_{0}=\beta_{0}\omega$ with $\beta_{0}=T_{0}^{-1}$ as before, and we note $\zer{2}=\frac{\pi^{2}}{6}$).
Using the expansion for $\bes{1}{-2}{0}$ of Appendix \ref{app:meb}, valid for small $x$ and small 
$\varepsilon x$ we obtain
\begin{equation}
\label{44170}
0=\left[1-\left(\frac{x}{x_{0}}\right)^{2}\right]\zer{2}
+\left(\varepsilon-\frac{1}{2}\right)x\ln{(\varepsilon x)}-\varepsilon x + \dots.
\end{equation}
Concentrating on the temperature $T=T_{0}$ at which $\varepsilon=\varepsilon_{0}$ we find to lowest order,
that $\varepsilon_{0}$ must satisfy
\begin{equation}
\label{44180}
0=\left(\varepsilon_{0}-\frac{1}{2}\right)\ln{\left(\varepsilon_{0} x_{0}\right)}-\varepsilon_{0}.
\end{equation}
Assuming $\varepsilon_{0}$ is finite as $x_{0}\rightarrow 0$, then in (\ref{44180}) we may use 
$|\ln{x_{0}}|\gg|\ln{\varepsilon_{0}}|$ to find
\begin{equation}
\label{44190}
\varepsilon_{0}\simeq\frac{1}{2}\left(1+\frac{1}{\ln{x_{0}}}\right)
\end{equation}
if only the leading term is kept. This shows that $\varepsilon_{0}\rightarrow\frac{1}{2}$ from below as 
$x_{0}\rightarrow 0$. Unlike the case $D=3$, for which $\lim_{B\rightarrow 0}{\frac{\mu}{B}}=\frac{1}{2}-a$
was non--zero, this time we find $\lim_{B\rightarrow 0}{\frac{\mu}{B}}=0$. Again by including more terms
it is possible to render the approximation for $\varepsilon_{0}$ more accurate. Because of the presence of the 
logarithm in (\ref{44190}), the convergence of $\varepsilon_{0}$ towards the value of $\frac{1}{2}$ is very slow.

If we expand the magnetization for $T=T_{0}$ using (\ref{44190}) it is found that 
\begin{equation}
\label{44200}
M\simeq - \frac{Q}{2mV}\frac{x_{0}}{\pi^{2}}\ln{\left(\frac{2}{x_{0}}\right)}
\end{equation}
for small $x_{0}$ (Only the leading term has been included here). For the specific heat we have 
\begin{equation}
\label{44210}
\frac{C_{v}}{N}\simeq 6\frac{\zer{3}}{\zer{2}}+ \frac{4\zer{2}}{\ln{\left(\frac{x_{0}}{2}\right)}}+\dots
\end{equation}
if only the leading order term is kept. Again we see the approach of the specific heat towards the free field
value at $T=T_{0}$ as $B\rightarrow 0$, and that the specific heat is lower when $B\neq 0$. The analytical 
results are consistent with the numerical ones presented earlier.

For $D\geq 5$ we may use
\begin{equation}
\label{44220}
x_{0} \bes{1}{2-D}{0} |_{x_{0}}=\zer{\frac{D}{2}}
\end{equation}
to solve for $\varepsilon_{0}$. (We still use $x_{0}=\beta_{0}\omega$ and $\varepsilon_{0}$ to denote the 
value of $\varepsilon$ at $T=T_{0}$). From our earlier discussion we know that $T_{0}>T_{\sss C}$ so that
$\varepsilon_{0}\neq 0$. The expansion for $\bes{1}{2-D}{0}$ depends upon whether $D$ is even or odd. For odd
$D$ we use (\ref{A1160}) with $\alpha=2-D$ and $\delta=0$ to obtain
\begin{eqnarray}
\nn
0&=&\zer{\frac{D-2}{2}}\left(\frac{1}{2}-\varepsilon_{0}\right) + \frac{1}{2}\zer{\frac{D-2}{2}}
\left(\varepsilon_{0}^{2}-\varepsilon_{0}+\frac{1}{6}\right) x_{0} \\
\label{44230}
&&+ \Gamma\left(\frac{4-D}{2}\right)\zeh{\frac{4-D}{2}}{\varepsilon_{0}}x_{0}^{\frac{D-4}{2}}
+{\cal O}\left(x_{0}^{2}\right).
\end{eqnarray}
By letting $x_{0}\rightarrow 0$ (corresponding to $B\rightarrow 0$) we see that 
$\varepsilon_{0}\rightarrow\frac{1}{2}$. In order to see how fast $\varepsilon_{0}\rightarrow\frac{1}{2}$
as $x_{0}\rightarrow 0$ we must distinguish between $D=5$ and $D>5$. For $D=5$ the next to leading order 
term is of order $x_{0}^{\frac{1}{2}}$ and we find
\begin{equation}
\label{44240}
\varepsilon_{0} = \frac{1}{2} + \frac{1}{2}\left(\frac{1}{\sqrt{2}}-1\right)
\left(\frac{x_{0}}{\pi}\right)^{\frac{1}{2}} + {\cal O}\left(x_{0}\right).
\end{equation}
For $D>5$ the next to leading order term is of order $x_{0}$ and we find
\begin{equation}
\label{44250}
\varepsilon_{0} = \frac{1}{2} - \frac{1}{24}\frac{\zer{\frac{D-4}{2}}}{\zer{\frac{D-2}{2}}} + \dots.
\end{equation}
(The next term in (\ref{44250}) is of order $x_{0}^{\frac{3}{2}}$ if $D=7$ and $x_{0}^{2}$ for $D>7$).

In the even dimensional case we may use (\ref{A1150}) with $k=\frac{D-2}{2}$ and $\delta=0$ in (\ref{44220}).
For $D=6$ we find
\begin{equation}
\label{44260}
\varepsilon_{0} = \frac{1}{2} + \frac{x_{0}\ln{x_{0}}}{4\pi^{2}} + {\cal O}\left(x_{0}\right).
\end{equation}
For $D>6$ the result in (\ref{44250}) holds. Therefore for all $D\geq 5$ we find that 
$\varepsilon_{0}\rightarrow \frac{1}{2}$ as $B\rightarrow 0$, proving that $\frac{\mu}{B}\rightarrow 0$
as $B\rightarrow 0$. This contrasts for the result for $D=3$.

\section{Discussion and Conclusions}
\label{sec:conc}

This paper has studied the thermodynamic properties of the ideal charged Bose gas in some detail. Spatial 
dimensions with $D\geq 3$ have been examined since even for the free gas it is known that the properties of 
the gas are sensitive to the spatial dimension. The specific heat was calculated numerically as well as 
analytically, for small values of the magnetic field. We also performed calculations of the magnetization
and showed how the effective action method could be used to account for the condensate when $D\geq 5$.

One motivation for our study was to understand in more detail the behavior of the magnetized gas for $D=3$.
When a magnetic field is present, no matter how small, there is no phase transition; however when there is 
no magnetic field the system does exhibit a phase transition with a non--zero condensate. On physical grounds 
we would expect that there should be some sense in which the system in a small magnetic field should
behave in almost the same way as the free gas. By examining the specific heat it is possible to see that as
the magnetic field is reduced, the curves start to resemble the specific heat for the free gas. So long as 
the magnetic field remains non--zero the specific heat is always smooth, with the specific heat maximum 
approaching the the free Bose gas transition temperature as the magnetic field is reduced. The results of
section \ref{sec:meb} may be used to study this analytically. Although we did not show it, it is
straightforward to show that the derivative of the specific heat becomes discontinuous as $B\rightarrow 0$,
exactly as in the case for the free Bose gas. Similar remarks apply to gases in other dimensions.

We also studied the behavior in large magnetic fields. Here we found that the specific heat for a gas in $D$
spatial dimensions looked like the specific heat for the free gas in $(D-2)$ spatial dimensions over a range 
of temperatures. Once the temperature gets too large this effective reduction in dimension disappears.

By using the Mellin--Barnes integral transform we were able to obtain a number of analytical approximations.
Although, for $D\geq 5$, it is not possible to solve for the critical temperature exactly, it is possible to
obtain good estimates when the magnetic field is weak. These approximations can then be used to study the
Meissner--Ochsenfeld effect. We were also able to obtain reliable approximations for the first time valid 
at the free gas condensation temperature. As a by--product our calculations showed exactly how the 
chemical potential vanishes as $B\rightarrow 0$.

The most obvious way that the calculations given in this paper should be extended is by the inclusion of 
Coulomb interactions among the charged particles. Clearly this is essential before it will be possible
to study Bose--Einstein condensation of charged particles in a reliable way. Given the recent experimental 
advances in the cooling and trapping of atomic gases, it may become increasingly important to study this 
problem for trapped ions. A less pressing extension of our work is to relativistic charged particles. Daicic
and Frankel \cite{daicfrangailkowa94a,daicfran96a} have done a study of this already in various cases;
however, it would be easily possible to extend the analysis of our paper to obtain the specific heat for the
first time.

\acknowledgments

GBS is grateful to the EPSRC for grant 94004194. DJT would like to thank K. Kirsten, F. Laloe, and S. Ouvry 
for helpful discussions. We are both grateful to J. Daicic and N. Frankel for helping us to clarify comments 
concerning Refs.~\cite{daicfran96a,ariajoan89a}.

\appendix
\section{Details of the Mellin--Barnes Transformations}
\label{app:meb}

\subsection{$\kappa=1$ case}

The sum $\bes{1}{\alpha}{\delta}$ defined in (\ref{32110}) is
\begin{equation}
\label{A1010}
\bes{1}{\alpha}{\delta} = \sum_{l=1}^{\infty} \frac{l^{\frac{\alpha}{2}} 
e^{-l x (\varepsilon+\delta)}}{\left(1-e^{-l x}\right)}.
\end{equation}
By using the binomial expansion on the denominator we find
\begin{equation}
\label{A1020}
\bes{1}{\alpha}{\delta} = \sum_{l=1}^{\infty} \sum_{n=0}^{\infty} l^{\frac{\alpha}{2}} 
e^{-l x (n+\varepsilon+\delta)}.
\end{equation}
This is now in a form where (\ref{41010}) can be used for the exponential. There are a variety of ways to use 
(\ref{41010}) on (\ref{A1020}), depending on how we split up the exponential. We require results for 
$\delta=0, 1$.

Suppose that we concentrate on $\delta=0$ initially. If we are interested in the $\varepsilon\rightarrow 0$ 
limit, it proves advantageous to separate off the $n=0$ term in (\ref{A1020}) before using (\ref{41010}). 
This is a slight variation on the method of reference \cite{daicfran96a} which is useful for isolating terms 
which diverge as $\varepsilon\rightarrow 0$, although we do not have to do this. We find
\begin{eqnarray}
\nn
\bes{1}{\alpha}{0} &=& \li{-\frac{\alpha}{2}}{e^{-x\varepsilon}} 
+ \int_{c-i\infty}^{c+i\infty} \frac{d\theta}{2\pi i}\Gamma(\theta) x^{-\theta} \\
\label{A1030}
&& \times \zeta_{\sss R} \left(\theta-\frac{\alpha}{2}\right) 
\zeta_{\sss H}\left(\theta,\varepsilon+1\right)
\end{eqnarray}
if it is noted that the sums over $l$ and $n$ can be done in terms of the Riemann and Hurwitz 
$\zeta$--functions respectively. If the $n=0$ term is not removed, then in place of (\ref{A1030}) we find 
just the integral part with $\zeh{\theta}{\varepsilon}$ in place of $\zeh{\theta}{1+\varepsilon}$. 
We take $c>0$ real and large enough so that the contour in (\ref{A1030}) lies to the right of the largest 
pole of the integrand (specifically, $c>\max\left(1,1+\frac{\alpha}{2}\right)$, where $\max$ denotes the 
maximum of the set). The contour in (\ref{A1030}) is closed in the left hand side of the complex plane, and 
the residue theorem used to generate the asymptotic expansion of $\bes{1}{\alpha}{0}$ from the known 
properties of the $\Gamma$-- and $\zeta$--functions \cite{whitwats02a}.

At this stage it is necessary to identify a number of special cases. If $\alpha=0$, then the poles of the 
Riemann and Hurwitz $\zeta$--functions both occur at $\theta=1$ and the integrand has a double pole. 
If $\alpha=-2 k$ where $k=1,2,3,\dots$, then the pole of the Riemann $\zeta$--function coincides with one 
of the poles of the $\Gamma$--function and again the integrand has a double pole. For 
$\alpha\neq 0,-2,-4,\dots$ the functions occurring in (\ref{A1030}) all have distinct poles, so that the 
integrand only has simple poles.

\subsubsection{$\alpha=0$, $\delta=0$}

Setting $\alpha=0$ in (\ref{A1030}) results in $\li{0}{e^{-\varepsilon x}}$. This is simply
\begin{equation}
\label{A1040}
\li{0}{e^{-\varepsilon x}}=\frac{1}{\left(e^{\varepsilon x}-1\right)}
\end{equation}
which follows directly from the definition (\ref{21040}). Evaluation of the integral by residues results in 
\begin{eqnarray}
\nn
\bes{1}{0}{0}&\simeq& \frac{1}{\left(e^{\varepsilon x}-1\right)} 
-\frac{\psi(1+\varepsilon)+\ln x}{x}  \\
\nn
&&- \sum_{p=0}^{\infty} \frac{(x)^{2p+1}}{\Gamma(2p+2)}\zer{-1-2p} \\
\label{A1050}
&& \times \zeh{-1-2p}{1+\varepsilon}.
\end{eqnarray}
We use the symbol $\simeq$ here to denote that the result is an asymptotic expansion. The Riemann and 
Hurwitz $\zeta$--functions in (\ref{A1050}) may be related to the Bernoulli numbers and polynomials 
\cite{whitwats02a}. The divergence as $\varepsilon\rightarrow 0$ is contained in the first term. When 
$\varepsilon x$ is small, we may use \cite{whitwats02a}
\begin{equation}
\label{A1060}
\frac{1}{\left(e^{\varepsilon x}-1\right)}=\frac{1}{\varepsilon x}-\frac{1}{2}-\sum_{p=0}^{\infty} 
\frac{(\varepsilon x)^{2p+1}}{\Gamma(2p+2)}\zer{-1-2p} 
\end{equation}
to obtain
\begin{eqnarray}
\nn
\bes{1}{0}{0}&\simeq& \frac{1}{\varepsilon x}-\frac{1}{2} 
-\frac{1}{x} \left(\psi(1+\varepsilon)+\ln x\right) \\
\nn
&&- \sum_{p=0}^{\infty} \frac{(x)^{2p+1}}{\Gamma(2p+2)}\zer{-1-2p} \\
\label{A1070}
&&\times \left\{\varepsilon^{2p+1}+ \zeh{-1-2p}{1+\varepsilon}\right\}.
\end{eqnarray}

\subsubsection{$\alpha=-2k$, $k=1,2,3,\dots$; $\delta=0$}

We set $\alpha=-2 k$ in (\ref{A1030}) and note that the integrand has simple poles at $\theta=1$ and 
$\theta=-p$ for $p\neq k-1$, and a double pole at $\theta=1-k$. It is straightforward to expand about 
these poles and use the residue theorem to obtain
\begin{eqnarray}
\nn
\bes{1}{-2k}{0} &\simeq& \li{k}{e^{-x \varepsilon}} + \frac{\zer{k+1}}{x}  \\
\nn
&& +\frac{(-x)^{k-1}}{\Gamma(k)} \left[{\zeta}_{\sss H}'(1-k,1+\varepsilon) \right. \\
\nn
&&\left. +(\gamma+\psi(k)-\ln{x})\zeta_{\sss H}(1-k,1+\varepsilon)\right] \\
\nn
&&+ \sum_{\sstack{p=0,}{p\neq k-1}}^{\infty} \frac{(-x)^{p}}{\Gamma(p+1)} \zeta_{\sss R}\left(k-p\right) \\
\label{A1080}
&& \times \zeh{-p}{1+\varepsilon}.
\end{eqnarray}
Here ${\zeta}_{\sss H}'(1-k,1+\varepsilon)$ denotes the derivative of $\zeh{s}{1+\varepsilon}$ with 
respect to $s$ evaluated at $s=1-k$. For $k=1$, ${\zeta}_{\sss H}'(0,1+\varepsilon)$ may be found in 
\cite{whitwats02a}. It is possible to generalize the procedure described in \cite{whitwats02a} to 
obtain ${\zeta}_{\sss H}'(1-k,1+\varepsilon)$ for other values of $k$ , although we have not found it possible 
to obtain as simple a result for $k=1$. The values we need are given in Appendix \ref{app:derhur}.

The polylogarithm $\li{k}{e^{-\varepsilon x}}$ may be expanded for small $\varepsilon x$, also using 
(\ref{41010}). This was originally described by Robinson \cite{robi51a}. We find 
\begin{eqnarray}
\nn
\li{k}{e^{-\varepsilon x}} &\simeq&\left(\frac{(-\varepsilon x)^{k-1}
\left(\psi(k)+\gamma-\ln{(\varepsilon x)}\right)}{\Gamma(k)}\right) \\
\label{A1090}
&& + \sum_{\sstack{l=0;}{l\neq k-1}}^{\infty} \frac{(-\varepsilon x)^{l}}{\Gamma(l+1)} \zeta_{\sss R}(k-l).
\end{eqnarray}
Finally we note that \cite{whitwats02a}
\begin{equation}
\label{A1100}
\psi(k)+\gamma = 1+\frac{1}{2}+\dots +\frac{1}{k-1}
\end{equation}
for $k>1$, and vanishes for $k=1$.

\subsubsection{$\alpha\neq 0,-2,-4,\dots$; $\delta=0$}

In this case all of the poles of the integrand in (\ref{A1030}) are simple, and we obtain
\begin{eqnarray}
\nn
\bes{1}{\alpha}{0} &\simeq& \li{-\frac{\alpha}{2}}{e^{-x \varepsilon}} 
+ \frac{\zeta_{\sss R}\left(1-\frac{\alpha}{2}\right)}{x} \\
\nn
&&+\Gamma\left(1+\frac{\alpha}{2}\right) x^{-\left(1+\frac{\alpha}{2}\right)} 
\zeta_{\sss H}\left(1+\frac{\alpha}{2},1+\varepsilon\right) \\
\nn
&&+ \sum_{p=0}^{\infty} \frac{(-x)^{p}}{\Gamma(p+1)}\zeta_{\sss R}  
\left(-\left[p+\frac{\alpha}{2}\right]\right) \\
\label{A1110}
&& \times \zeh{-p}{1+\varepsilon}.
\end{eqnarray}
The expansion of the polylogarithm for small $\varepsilon x$ is again described in reference 
\cite{robi51a} and is
\begin{eqnarray}
\nn
\li{-\frac{\alpha}{2}}{e^{-\varepsilon x}} &\simeq& \Gamma(1+\frac{\alpha}{2}) 
(\varepsilon x)^{-1-\frac{\alpha}{2}} \\
\label{A1120}
&&+ \sum_{l=0}^{\infty} \frac{(-\varepsilon x)^{l}}{\Gamma(l+1)} \zeta_{\sss R}(-l-\frac{\alpha}{2}).
\end{eqnarray}
\medskip
We also require the expansion of $\bes{1}{\alpha}{\delta}$ for $\delta>0$. The difference between 
$\delta=0$ and $\delta \neq 0$ can be seen from (\ref{A1020}): when $\delta=0$ the $n=0$ term can lead 
to a divergent sum over $l$ when $\varepsilon\rightarrow 0$. For $\delta>0$ the sum over $l$ is always 
convergent, even when $\varepsilon=0$. Thus for $\delta>0$ we do not have to separate off the $n=0$ term . 
Using (\ref{41010}) for (\ref{A1020}) we obtain
\begin{equation}
\label{A1130}
\bes{1}{\alpha}{\delta} =  \int_{c-i\infty}^{c+i\infty} \frac{d\theta}{2\pi i}\Gamma(\theta) x^{-\theta} 
\zeta_{\sss R} \left(\theta-\frac{\alpha}{2}\right) \zeta_{\sss H}\left(\theta,\varepsilon+\delta\right).
\end{equation}
The same cases arise as for $\delta=0$, and the analysis is similar. We will simply list the results.

\subsubsection{$\alpha=0$, $\delta>0$}

\begin{eqnarray}
\nn
\bes{1}{0}{\delta}&\simeq& -\frac{1}{x} \left(\psi(\varepsilon+\delta)+\ln x\right) \\
\nn
&&- \sum_{p=0}^{\infty} \frac{x^{2p+1}}{\Gamma(2p+2)}\zer{-1-2p} \\
\label{A1140}
&&\times \zeh{-1-2p}{\varepsilon+\delta}.
\end{eqnarray}

\subsubsection{$\alpha=-2k$, $k=1,2,3,\dots$; $\delta>0$}

\begin{eqnarray}
\nn
\bes{1}{-2k}{\delta} &\simeq& \frac{\zer{k+1}}{x} \\
\nn
&&+\frac{(-x)^{k-1}}{\Gamma(k)} \left[{\zeta}_{\sss H}'(1-k,\varepsilon+\delta) \right.\\
\nn
&&\left. +(\gamma+\psi(k)-\ln{x})\zeta_{\sss H}(1-k,\varepsilon+\delta)\right] \\
\nn
&&+ \sum_{\sstack{p=0,}{p\neq k-1}}^{\infty} \frac{(-x)^{p}}{\Gamma(p+1)}\zeta_{\sss R}\left(k-p\right)  \\
\label{A1150}
&&\times \zeh{-p}{\varepsilon+\delta}.
\end{eqnarray}

\subsubsection{$\alpha\neq 0,-2,-4,\dots$; $\delta>0$}

\begin{eqnarray}
\nn
\bes{1}{\alpha}{\delta} &\simeq& \frac{\zeta_{\sss R}\left(1-\frac{\alpha}{2}\right)}{x} \\
\nn
&&+\Gamma\left(1+\frac{\alpha}{2}\right) x^{-\left(1+\frac{\alpha}{2}\right)} 
\zeta_{\sss H}\left(1+\frac{\alpha}{2},\varepsilon+\delta\right) \\
\nn
&&+ \sum_{p=0}^{\infty} \frac{(-x)^{p}}{\Gamma(p+1)}\zeta_{\sss R}  
\left(-\left[p+\frac{\alpha}{2}\right]\right) \\
\label{A1160}
&& \times \zeh{-p}{\varepsilon+\delta}.
\end{eqnarray}

One final comment we wish to make is that (\ref{A1140}---\ref{A1160}) hold even for 
$\delta=0$, although as we explained, in this case it may be more difficult to isolate the terms which 
diverge as $\varepsilon\rightarrow 0$. Nevertheless, these results can be used if desired.

\subsection{$\kappa=2$ case}

The sum $\bes{2}{\alpha}{\delta}$ defined in (\ref{32110}) is
\begin{equation}
\label{A2010}
\bes{2}{\alpha}{\delta} = \sum_{l=1}^{\infty} \frac{l^{\frac{\alpha}{2}} 
e^{-l x (\varepsilon+\delta)}}{\left(1-e^{-l x}\right)^{2}}.
\end{equation}
The analysis of this sum proceeds in the same way as for $\bes{1}{\alpha}{\delta}$. Binomial expansion of 
the denominator of (\ref{A2010}) gives
\begin{equation}
\label{A2020}
\bes{2}{\alpha}{\delta} = \sum_{l=1}^{\infty} \sum_{n=0}^{\infty} l^{\frac{\alpha}{2}} (n+1) 
e^{-l x (n+\varepsilon+\delta)}.
\end{equation}
We only require the result for $\delta=1$ in the main part of the paper, so that the sum over $l$ is 
convergent even for $\varepsilon=0$. We do not separate off any terms. Application of (\ref{41010}) to 
(\ref{A2020}) results in
\begin{eqnarray}
\nn
\bes{2}{\alpha}{\delta}&=&  \int_{c-i\infty}^{c+i\infty} \frac{d\theta}{2\pi i} \Gamma(\theta) x^{-\theta} 
\zeta_{\sss R} \left(\theta-\frac{\alpha}{2}\right) \\
\nn
&& \times \left\{\zeta_{\sss H}\left(\theta-1,\varepsilon+\delta\right) \right. \\
\label{A2030}
&& \left. +(1-\varepsilon-\delta)\zeta_{\sss H}\left(\theta,\varepsilon+\delta\right)\right\}.
\end{eqnarray}
(The sum over $l$ is a Riemann $\zeta$--function, and the sum over $n$ can be done in terms of the 
Hurwitz $\zeta$--function as shown).

\subsubsection{$\alpha=0$}

The integrand of (\ref{A2030}) has a simple pole at $\theta=2$, a double pole at $\theta=1$, and simple poles 
at $\theta=0,-1,-2,\dots$. We find
\begin{eqnarray}
\nn
\bes{2}{0}{\delta} &\simeq& \frac{\zeta_{\sss R}(2)}{x^{2}} -\frac{1}{4}(\varepsilon+\delta)^{2}
+\frac{1}{2}(\varepsilon+\delta) - \frac{5}{24}\\
\nn
&&+\frac{1}{x}\left[\frac{1}{2}-\varepsilon-\delta-(1-\varepsilon-\delta)\psi(\varepsilon+\delta) \right. \\
\nn
&&\left. -(1-\varepsilon-\delta)\ln x\right] \\
\nn
&&+\sum_{p=0}^{\infty} \frac{(x)^{2p+1}}{\Gamma(2p+2)} \zeta_{\sss R} \left(-1-2p\right)  \\
\nn
&&\times \left\{\zeh{-2-2p}{\varepsilon+\delta} \right. \\
\label{A2040}
&& \left. +(1-\varepsilon-\delta)\zeh{-1-2p}{\varepsilon+\delta}\right\}.
\end{eqnarray}

\subsubsection{$\alpha=2$}

The integrand of (\ref{A2030}) has a double pole at $\theta=2$, and single poles at $\theta=1,0,-1,\dots$. 
The residue theorem gives,
\begin{eqnarray}
\nn
\bes{2}{2}{\delta} &\simeq& -\frac{(1-\varepsilon-\delta)}{2x} \\
\nn
&&+\frac{1}{x^{2}}\left[1-\ln{x}-\psi(\varepsilon+\delta)\right. \\
\nn
&&\left. +(1-\varepsilon-\delta)\zeh{2}{\varepsilon+\delta}\right] \\
\nn
&&+\sum_{p=0}^{\infty} \frac{(x)^{2p}}{\Gamma(2p+1)} \zeta_{\sss R} \left(-1-2p\right)  \\
\nn
&&\times \left\{\zeh{-1-2p}{\varepsilon+\delta} \right. \\
\label{A2050}
&& \left. +(1-\varepsilon-\delta)\zeh{-2p}{\varepsilon+\delta}\right\}.
\end{eqnarray}

\subsubsection{$\alpha=-2k$, $k=1,2,3,\dots$}

The integrand of (\ref{A2030}) has simple poles at $\theta=2, 1$, a double pole at $\theta=1-k$, and simple 
pole at $\theta=-p$, $p=0,1,2,3,\dots$ with $p\neq k-1$. We find
\begin{eqnarray}
\nn
\bes{2}{-2k}{\delta}&=& \frac{\zeta_{\sss R}\left(2+k\right)}{x^{2}} + \frac{(1-\varepsilon-\delta)
\zeta_{\sss R}\left(1+k\right)}{x} \\
\nn
&&+\frac{(-x)^{k-1}}{\Gamma(k)}(\gamma+\psi(k)-\ln{x}) \\
\nn
&&\times\left\{\zeta_{\sss H}\left(-k,\varepsilon+\delta\right) \right.\\
\nn
&&\left. +(1-\varepsilon-\delta)\zeta_{\sss H}\left(1-k,\varepsilon+\delta\right)\right\} \\
\nn
&&+ \frac{(-x)^{(k-1)}}{\Gamma(k)} \left\{\zeta_{\sss H}'\left(-k,\varepsilon+\delta\right) \right. \\
\nn
&&\left. +(1-\varepsilon-\delta)\zeta_{\sss H}'\left(1-k,\varepsilon+\delta\right)\right\} \\
\nn
&&+\sum_{\sstack{p=0,}{p\neq k-1}}^{\infty} \frac{(-x)^{p}}{\Gamma(p+1)} \zeta_{\sss R} (k-p)  \\
\nn
&& \times \left\{\zeh{-p-1}{\varepsilon+\delta} \right. \\
\label{A2060}
&& \left. +(1-\varepsilon-\delta)\zeh{-p}{\varepsilon+\delta}\right\}.
\end{eqnarray}

\subsubsection{$\alpha\neq 2,0,-2,-4,\dots$}

The integrand only has simple poles in this case, and we find that
\begin{eqnarray}
\nn
\bes{2}{\alpha}{\delta}&=& \Gamma\left(1+\frac{\alpha}{2}\right) x^{-\left(1+\frac{\alpha}{2}\right)} 
\left\{\zeta_{\sss H}\left(\frac{\alpha}{2},\varepsilon+\delta\right) \right. \\
\nn
&&\left. +(1-\varepsilon-\delta)\zeta_{\sss H}\left(1+\frac{\alpha}{2},\varepsilon+\delta\right)\right\} \\
\nn
&& + \frac{\zeta_{\sss R}\left(2-\frac{\alpha}{2}\right)}{x^{2}} 
+ \frac{(1-\varepsilon-\delta)\zeta_{\sss R}\left(1-\frac{\alpha}{2}\right)}{x} \\
\nn
&&+\sum_{p=0}^{\infty} \frac{(-x)^{p}}{\Gamma(p+1)} \zeta_{\sss R} \left(-p-\frac{\alpha}{2}\right)  \\
\nn
&& \times \left\{\zeh{-p-1}{\varepsilon+\delta} \right. \\
\label{A2070}
&& \left. +(1-\varepsilon-\delta)\zeh{-p}{\varepsilon+\delta} \right\}.
\end{eqnarray}

\subsubsection{$\kappa=3$ case}

From (\ref{32110}) we have
\begin{equation}
\label{A3010}
\bes{3}{\alpha}{\delta} = \sum_{l=1}^{\infty} \frac{l^{\frac{\alpha}{2}} 
e^{-l x (\varepsilon+\delta)}}{\left(1-e^{-l x} \right)^{3}}.
\end{equation}
Again we are only interested in $\delta>0$. Following the now familiar steps of binomially expanding the 
denominator, using (\ref{41010}) for the exponential, and performing the sums in terms of $\zeta$--functions 
results in
\begin{eqnarray}
\nn
\bes{3}{\alpha}{\delta} &\simeq& \int_{c-i\infty}^{c+i\infty} \frac{d\theta}{2\pi i} \Gamma(\theta) 
\frac{x^{-\theta}}{2} \zeta_{\sss R} \left(\theta-\frac{\alpha}{2}\right) \\
\nn
&& \times \left\{\zeta_{\sss H} \left(\theta-2,\varepsilon+\delta\right) \right. \\
\nn
&&+ (3-2\varepsilon-2\delta)\zeta_{\sss H} \left(\theta-1,\varepsilon+\delta\right) \\
\label{A3020}
&& \left. + (1-\varepsilon-\delta)(2-\varepsilon-\delta)
\zeta_{\sss H} \left(\theta,\varepsilon+\delta\right)\right\}.
\end{eqnarray}

\subsubsection{$\alpha=0$}

The integrand in (\ref{A3020}) has simple poles at $\theta=3, 2, 0, -1, -2, \dots$ and a double pole at 
$\theta=1$. We find 
\begin{eqnarray}
\nn
\bes{3}{0}{\delta}&=& \frac{\zeta_{\sss R} (3)}{x^{3}} 
+\frac{(3-2\varepsilon-2\delta)\zeta_{\sss R}(2)}{2x^{2}} \\
\nn
&&+\frac{1}{2x}\left[\frac{3}{2}(\varepsilon+\delta)^{2}-\frac{7}{2}(\varepsilon+\delta)+\frac{17}{12} \right. \\
\nn
&&\left. -(1-\varepsilon-\delta)(2-\varepsilon-\delta)\left\{\psi(\varepsilon+\delta)+\ln x\right\}\right] \\
\nn
&& - \frac{1}{4}\left\{\zeta_{\sss H} \left(-2,\varepsilon+\delta\right) \right. \\
\nn
&& + (3-2\varepsilon-2\delta)\zeta_{\sss H} \left(-1,\varepsilon+\delta\right) \\
\nn
&& \left. +(1-\varepsilon-\delta)(2-\varepsilon-\delta)\left(\frac{1}{2}-\varepsilon-\delta \right)\right\} \\
\nn
&-& \sum_{p=0}^{\infty} \frac{(x)^{2p+1}}{2\Gamma(2p+2)} \zeta_{\sss R} (-1-2p) \\
\nn
&&\times \left\{\zeh{-3-2p}{\varepsilon+\delta} \right. \\
\nn
&&\left. + (3-2\varepsilon-2\delta)\zeh{-2-2p}{\varepsilon+\delta} \right. \\
\nn
&& +(1-\varepsilon-\delta)(2-\varepsilon-\delta) \\
\label{A3030}
&&\left. \times \zeh{-1-2p}{\varepsilon+\delta}\right\}.
\end{eqnarray}

\subsubsection{$\alpha=2, 4$}

The integrand of (\ref{A3020}) has simple poles at $\theta=3, 1, 0, -1, -2, \dots$ and a double pole at 
$\theta=2$ for $\alpha=2$. When $\alpha=4$ we have a double pole at $\theta=3$, and simple poles at 
$\theta=2, 1, 0, \dots$. Because we do not require these cases in our paper we will not give the results 
here.

\subsubsection{$\alpha=-2k$, $k=1,2,3,\dots$}

The integrand of (\ref{A3020}) has simple poles at $\theta=3, 2, 1, -p$ for $p=0, 1, 2, \dots$ but 
$p\neq k-1$, and a double pole at $\theta=1-k$. We find
\begin{eqnarray}
\nn
\bes{3}{-2k}{\delta}&\simeq& \frac{\zeta_{\sss R}(3+k)}{x^{3}} \\
\nn
&&+\frac{(3-2\varepsilon-2\delta)\zeta_{\sss R}(2+k)}{2x^{2}}\\
\nn
&& +\frac{(1-\varepsilon-\delta)(2-\varepsilon-\delta)\zeta_{\sss R} (1+k)}{2x}\\
\nn
&& +\frac{(-x)^{k-1}}{2\Gamma(k)}\left\{[\gamma+\psi(k)] \right.\\
\nn
&&\left[\zeta_{\sss H} \left(-1-k,\varepsilon+\delta\right) \right. \\
\nn
&& + (3-2\varepsilon-2\delta)\zeta_{\sss H} \left(-k,\varepsilon+\delta\right) \\
\nn
&&\left. + (1-\varepsilon-\delta)(2-\varepsilon-\delta)\zeta_{\sss H} \left(1-k,\varepsilon+\delta\right) 
\right] \\
\nn
&& + \zeta_{\sss H}' \left(-1-k,\varepsilon+\delta\right)  \\
\nn
&&+ (3-2\varepsilon-2\delta)\zeta_{\sss H}' \left(-k,\varepsilon+\delta\right) -\ln x \\
\nn
&& \left. + (1-\varepsilon-\delta)(2-\varepsilon-\delta)\zeta_{\sss H}' \left(1-k,\varepsilon+\delta\right)
\right\} \\
\nn
&&+ \sum_{\sstack{p=0,}{p\neq 1-k}}^{\infty} \frac{(-x)^{p}}{2\Gamma(p+1)} \zeta_{\sss R}(k-p) \\
\nn
&&\times \left\{\zeh{-p-2}{\varepsilon+\delta} \right. \\
\nn
&&\left. + (3-2\varepsilon-2\delta)\zeh{-p-1}{\varepsilon+\delta} 
\right.\\
\label{A3040}
&&\left. + (1-\varepsilon-\delta)(2-\varepsilon-\delta)\zeh{-p}{\varepsilon+\delta}\right\}.
\end{eqnarray}

\subsubsection{$\alpha\neq 4-2k$, $k=0, 1, 2, \dots$}

In this case all the poles of (\ref{A3020}) are simple and we obtain
\begin{eqnarray}
\nn
\bes{3}{\alpha}{\delta}&\simeq& \Gamma\left(1+\frac{\alpha}{2}\right) 
\frac{x^{-\left(1+\frac{\alpha}{2}\right)}}{2}\left\{ 
\zeta_{\sss H}\left(\frac{\alpha}{2}-1,\varepsilon+\delta\right) \right. \\
\nn
&&+(3-2\varepsilon-2\delta)\zeta_{\sss H}\left(\frac{\alpha}{2},\varepsilon+\delta\right) \\
\nn
&&\left. + (1-\varepsilon-\delta)(2-\varepsilon-\delta) \zeta_{\sss H} 
\left(1+\frac{\alpha}{2},\varepsilon+\delta\right)\right\} \\
\nn
&&+\frac{\zeta_{\sss R}\left(3-\frac{\alpha}{2}\right)}{x^{3}} 
+\frac{(3-2\varepsilon-2\delta)\zeta_{\sss R}\left(2-\frac{\alpha}{2}\right)}{2x^{2}} \\
\nn
&&+\frac{(1-\varepsilon-\delta)(2-\varepsilon-\delta)\zeta_{\sss R}\left(1-\frac{\alpha}{2}\right)}{2x} \\
\nn
&+& \sum_{p=0}^{\infty} \frac{(-x)^{p}}{2\Gamma(p+1)} 
\zeta_{\sss R}\left(-p+\frac{\alpha}{2}\right) \\
\nn
&& \times \left\{\zeh{-p-2}{\varepsilon+\delta} \right. \\
\nn
&&+ (3-2\varepsilon-2\delta)\zeh{-p-1}{\varepsilon+\delta} \\
\label{A4010}
&&\left. + (1-\varepsilon-\delta)(2-\varepsilon-\delta)\zeh{-p}{\varepsilon+\delta}\right\}.
\end{eqnarray}

\section{Derivatives of the Hurwitz $\zeta$--function}
\label{app:derhur}

The derivative of $\zeh{s}{a}$ with respect to $s$ at $s=0$ is given by \cite{whitwats02a}
\begin{equation}
\label{B1010}
\zeta_{\sss H}'(0,a)=\ln\left(\frac{\Gamma(a)}{\sqrt{2\pi}}\right).
\end{equation}
The derivative makes use of the Plana summation formula to derive an integral representation for 
$\zeh{s}{a}$ (the Hermite representation) which is then differentiated with respect to $s$. In  a similar 
manner the  derivatives at other values of $s$ may be calculated, although the results are less simple than 
(\ref{B1010}). For $s=-1$ we find
\begin{eqnarray}
\nn
\zeta_{\sss H}'(-1,a)&=&\frac{\left(6a^{2}(\gamma+1)+6a+(1-\gamma-\ln{2\pi})\right)}{12} \\
\nn
&&+\ln\left(\left\{\frac{\Gamma(a)}{\sqrt{2\pi}}\right\}^{a}\right) +\frac{\zeta_{\sss R}'(2)}{2\pi^{2}} \\
\label{B1020}
&&-\sum_{t=1}^{\infty}\frac{(-1)^{t+1}\zer{t+1}a^{(t+2)}}{(t+2)}.
\end{eqnarray}
Here $\zeta_{\sss R}'(2)$ is the derivative of the Riemann $\zeta$--function given by
\begin{equation}
\label{B1030}
\zeta_{\sss R}'(2) = -\sum_{n=1}^{\infty}\frac{\ln{n}}{n^{2}}
\end{equation}
which can be evaluated easily numerically. For $s=-2$ we find
\begin{eqnarray}
\nn
\zeta_{\sss H}'(-2,a)&=&\frac{\left(a^{3}(8\gamma+6)+9a^{2}+a(3-2\gamma-2\ln{2\pi})\right)}{12} \\
\nn
&&+\ln\left(\left\{\frac{\Gamma(a)}{\sqrt{2\pi}}\right\}^{a^{2}}\right)+\left(\frac{4a\zeta_{\sss R}'(2)
-\zer{3}}{4\pi^{2}}\right) \\
\label{B1040}
&&-\sum_{t=1}^{\infty}\frac{(-1)^{(t+1)}a^{(t+3)}(t+4)\zer{t+1}}{(t+2)(t+3)}.
\end{eqnarray}
A useful check on the results is provided by the general relation
\begin{equation}
\label{B1050}
\frac{\partial}{\partial a}\zeta_{\sss H}'(s,a)= -\zeh{s+1}{a}-s \zeta_{\sss H}'(s+1,a).
\end{equation}

\end{document}